\documentclass[a4paper]{article}
\usepackage[german, english]{babel}
\usepackage{a4wide}
\usepackage{amsmath}
\usepackage{amssymb}
\usepackage{ifthen}
\usepackage{epsfig}
\usepackage[hang,nooneline]{subfigure}
\newcounter{fig}   

\newcommand{\vphi}{\varphi}

\begin{document}

\title{\bf  Rotating Boson Stars in $\mathbf{5}$ Dimensions}
\vspace{1.5truecm}
\author{
{\bf Betti Hartmann$^1$, Burkhard Kleihaus$^2$, Jutta Kunz$^2$, 
and Meike List$^{3,2}$}\\[10pt]
$^1$ School of Engineering and Science, Jacobs University Bremen\\
D-28759 Bremen, Germany\\
\hspace{0.1cm}\\
$^2$ Institut f\"ur  Physik, Universit\"at Oldenburg, Postfach 2503\\
D-26111 Oldenburg, Germany\\
\hspace{0.1cm}\\
$^3$ ZARM, Universit\"at Bremen, Am Fallturm\\
D-28359 Bremen, Germany
}

\vspace{1.5truecm}

\date{\today}

\maketitle
\vspace{1.0truecm}

\begin{abstract}
We study rotating boson stars in five spacetime dimensions.
The boson fields consist of a complex doublet scalar field.
Considering boson stars rotating in two orthogonal planes with 
both angular momenta of equal magnitude, a special ansatz for the 
boson field and the metric allows for solutions with nontrivial 
dependence on the radial coordinate only.
The charge of the scalar field equals the
sum of the angular momenta. The rotating boson stars are 
globally regular and asymptotically flat. For our choice of a
sixtic potential the rotating boson star solutions possess a
flat spacetime limit. We study the solutions in flat and curved spacetime.
\end{abstract}

\section{Introduction}

Boson stars are globally regular configurations 
of self-gravitating boson fields
\cite{Lee:1991ax,Jetzer:1991jr,Mielke:2000mh,Schunck:2003kk}.
When the complex scalar field
has a suitable self-interaction,
the boson stars possess a flat spacetime limit,
representing non-topological solitons or $Q$-balls,
\cite{Friedberg:1976me,Coleman:1985ki}.
These are stationary localized solutions with a finite mass,
whose charge $Q$ corresponds to their particle number.
Supersymmetric extensions of the
Standard Model also possess $Q$-ball solutions \cite{Kusenko:1997ad}. 
In this case several scalar fields
interact via complicated potentials. 
It was shown that cubic interaction terms that result from
Yukawa couplings in the superpotential 
and supersymmetry breaking terms lead to the existence of $Q$-balls
with non-vanishing baryon or lepton number or electric charge. 
These supersymmetric $Q$-balls have been considered 
recently as possible candidates for baryonic dark matter 
\cite{Kusenko:2000fc} 
and their astrophysical implications have been discussed 
\cite{implications}.
Supersymmetric $Q$-balls have been constructed numerically in
\cite{Campanelli}.

The complex scalar field of boson stars
and $Q$-balls has a harmonic time-dependence
with frequency $\omega_s$.
In 4-dimensional flat spacetime,
the allowed frequency range for $Q$-balls
$\omega_{\rm min} < \omega_s < \omega_{\rm max}$
is solely determined by the properties of the potential
describing the self-interaction
\cite{Friedberg:1976me,Coleman:1985ki}.
This frequency range holds for both non-rotating
and rotating $Q$-balls 
\cite{Volkov:2002aj,Kleihaus:2005me,Kleihaus:2007vk}.
For 4-dimensional boson stars the maximal
frequency $\omega_{\rm max}$ remains the same as in flat spacetime,
whereas the minimal frequency depends on the strength
of the gravitational coupling constant and increases with increasing
gravitational coupling.

For $Q$-balls in 4 dimensions, the mass and charge diverge
at both ends of the frequency interval and assume
their minimal value at a critical frequency $\omega_{\rm cr}$,
where the classical stability of the solutions changes
\cite{Friedberg:1976me,Coleman:1985ki}.
For $\omega_s < \omega_{\rm cr}$,
i.e., below the critical value, the solutions are classically stable.
As long as their mass is smaller than the mass of $Q$ free bosons
they are also quantum mechanically stable
\cite{Friedberg:1976me}.
In constrast, above $\omega_{\rm cr}$ the solutions are unstable.

The coupling to gravity modifies and enriches this simple pattern.
Thus when $\omega_s \to \omega_{\rm max}$, 
the mass and charge 
of boson stars tend to zero, when the spacetime is 4-dimensional.
Moreover, in the lower frequency range, the boson star
solutions are not uniquely determined by their frequency.
Instead a spiral-like frequency dependence of the charge and the mass
is observed, where the charge and mass approach finite limiting values
at the center of the spiral.
Thus a plethora of boson star branches appears, typically featuring
a single classically stable branch
\cite{Kusmartsev:2008py}.

Remarkably, for charged boson fields with a $V$-shaped scalar potential,
also boson shells can arise \cite{Arodz:2008nm,Kleihaus:2009kr}.
These may harbour black holes, violating black hole uniqueness
\cite{Kleihaus:2010ep}.

The presence of rotation leads to an interesting feature
of these solutions,
namely their angular momentum is quantized
and corresponds to an integer $N$ times the charge,
$J=NQ$ \cite{Volkov:2002aj,Schunck:1996}.
Rotating $Q$-balls and boson stars with $N=1$ 
thus have the smallest angular momentum $J$
for a given charge $Q$.
While non-rotating $Q$-balls and boson stars
are cohomogeneity-1, the presence of rotation
includes an angular dependence
and thus leads to a cohomogeneity-2 problem in 4 dimensions.

With the advent of string theory the exploration of higher
dimensions has yielded many new insights for localized objects
and their properties. For black holes, in particular,
it became clear, that many features of black holes
in 4 dimensions do not extend to higher dimensions
\cite{Emparan:2008eg}.
It therefore appears eligible, to consider also other
localized objects in higher dimensions, such as
boson stars, which still represent alternatives to
black holes \cite{Cardoso:2007az}.
A first effort in this direction was undertaken
by Astefanesei and Radu \cite{Astefanesei:2003qy},
who addressed non-rotating boson stars with a negative cosmological
constant in 5 dimensions, and by Prikas \cite{Prikas:2004fx} who studied
electrically charged non-rotating boson stars in various dimensions in asymptotically
flat as well as Anti de-Sitter spacetime. 
Here we consider rotating asymptotically flat boson stars
and their flat spacetime counterparts.

In general, rotating objects in $D$ dimensions possess
$[(D-1)/2]$ independent angular momenta \cite{Myers:1986un}.
Thus in 5 dimensions, there are two independent angular
momenta associated with the two orthogonal planes of rotation.
Generic rotating solutions in 5 dimensions then depend on two coordinates,
just as in 4 dimensions.
However, when both angular momenta are chosen equal, $J_1=J_2=J$,
the symmetry of the solutions may be enhanced.
For black holes one then obtains a cohomogeneity-1 problem,
with dependence only on the radial coordinate,
also in the presence of Abelian gauge fields and dilaton fields
\cite{Kunz:2006eh}.

Here we show, that with an appropriate Ansatz for the
scalar fields, the problem can be reduced to cohomogeneity-1
also for rotating $Q$-balls and boson stars in 5 dimensions,
when both angular momenta are chosen equal.
For that purpose we introduce a complex doublet of scalar fields
as present, for instance, in the electroweak sector of the
standard model. 
Moreover, we show that the quantization of the angular momentum
also carries over, yielding for the lowest such rotating solutions
the relation $2|J|=Q$.

The paper is organized as follows.
In section 2 we present the action, the ansatz and the expressions
for the physical properties of the rotating $Q$-balls and boson stars
in 5 dimensions.
In section 3 we exhibit the numerical results for the solutions
in flat spacetime, comparing non-rotating and rotating $Q$-ball solutions,
and discussing their stability.
The effects of the coupling to gravity is the topic of section 4,
where we present the non-rotating and rotating boson star solutions.
We give our conclusions in section 5.

\section{Action, Ansatz and Charges}

\subsection{Action}

In 4 dimensions, $Q$-balls and boson stars with 2 complex scalar fields
have been considered before, leading to interesting
phenomena due to their interaction
\cite{Brihaye:2007tn}.
Here we introduce a complex doublet scalar field $\Phi$,
in order to obtain rotating cohomogeneity-1 
$Q$-balls and boson stars in 5 dimensions.

Thus we consider the following action of the self-interacting 
complex doublet scalar field $\Phi$ 
coupled minimally to Einstein gravity in 5 spacetime dimensions
\begin{equation}
S=\int \left[ \frac{R}{16\pi G}
   -  \left( \partial_\mu \Phi \right)^\dagger \left( \partial^\mu \Phi \right)
 - U( \left| \Phi \right|) 
 \right] \sqrt{-g} d^5x
 , \label{action}
\end{equation}
with
curvature scalar $R$ and Newton's constant 
(in 5 dimensions) $G$ and self-interaction potential $U$. 
Here $^\dagger$ denotes the complex transpose,
and $|\Phi|^2 = \Phi^\dagger \Phi $.

The scalar potential $U$ is chosen as
\begin{equation}
U(|\Phi|) =  \lambda \left( |\Phi|^6 -a |\Phi|^4 + b  |\Phi|^2 \right) 
 , \label{U} \end{equation} 
where $\lambda$, $a$, $b$ are constants.
The choice of potential allows for the existence of
nontopological soliton solutions,
i.e., $Q$-balls, in the absence of gravity.
This self-interaction of the scalar field has an attractive component,
and the potential has a minimum, $U(0)=0$, at $\Phi =0$
and a second minimum at some finite value of $|\Phi|$.
The boson mass is determined by the quadratic term and given
by $m_{\rm B}=\sqrt{\lambda b}$.

Variation of the action with respect to the metric and the scalar fields
leads, respectively, to the Einstein equations
\begin{equation}
G_{\mu\nu}= R_{\mu\nu}-\frac{1}{2}g_{\mu\nu}R = 8\pi G T_{\mu\nu}
\  \label{ee} \end{equation}
with stress-energy tensor
\begin{eqnarray}
T_{\mu\nu} &=& g_{\mu\nu}{L}_M
-2 \frac{\partial {L}_M}{\partial g^{\mu\nu}}
\nonumber\\
&=& -
   \frac{1}{2} 
   g_{\mu\nu} \left((\partial_\alpha \Phi)^\dagger (\partial_\beta \Phi)
  + (\partial_\beta \Phi)^\dagger (\partial_\alpha \Phi)\right) g^{\alpha\beta}
  + (\partial_\mu \Phi)^\dagger (\partial_\nu \Phi) 
  + (\partial_\nu \Phi)^\dagger (\partial_\mu \Phi)
\nonumber\\[5pt]
&&- 
    {\lambda} g_{\mu\nu}  U(|\Phi|) 
 , \label{tmunu}
\end{eqnarray}
and the matter field equation,
\begin{eqnarray}
& &\frac{1}{\sqrt{-g}}
\partial_\mu\left(\sqrt{-g} \partial^\mu \Phi \right) = 
 \frac{\partial U}{\partial |\Phi|^2} \Phi
 . \label{feqH} 
 \end{eqnarray}

\subsection{Ansatz}

To construct stationary rotating solutions with two equal angular momenta
we employ bi-azimuthal isotropic coordinates 
\cite{Kunz:2005nm,Kunz:2006eh}
\begin{eqnarray}
ds^2=g_{\mu\nu}dx^\mu dx^\nu=
& & 
-fdt^2 +\frac{m}{f}\left(dr^2+r^2 d\theta^2\right)
+\frac{m-n}{f}r^2 \sin^2\theta\cos^2\theta\left(d\vphi_1-d\vphi_2\right)^2                    
 \nonumber \\
& &
+\frac{n}{f}r^2 \left( \sin^2\theta (d\vphi_1-\frac{\omega}{r}dt)^2
                      +\cos^2\theta (d\vphi_2-\frac{\omega}{r}dt)^2 \right) \ . 
\label{metric_ansatz}
 \end{eqnarray}
This metric has 3 commuting Killing vector fields
\begin{equation}
\xi=\partial_t \ , \ \ \ \eta_1=\partial_{\varphi_{1}}
, \ \ \ \eta_2=\partial_{\varphi_2}
\ , \label{xieta} \end{equation}
where $\xi$ denotes an asymptotically timelike
Killing vector field and $\eta_{i}$ denote asymptotically spacelike
Killing vector fields.
The 4 metric functions
$f$, $m$, $n$ and $\omega$ are functions of 
the radial variable $r$ only.
In the non-rotating case, the metric simplifies,
since $m=n$ and $\omega=0$.
In the flat spacetime limit, all metric functions become trivial,
$f=m=n=1$, $\omega=0$.

For the complex doublet boson field we assume a harmonic
time-dependence with frequency $\omega_s$.
The Ansatz then has the form
\begin{equation}
 \Phi = \phi(r) e^{i \omega_s t} \hat{\Phi}
 , \label{phi} 
\end{equation}
where $\hat{\Phi}$ denotes a complex unit two vector.
To specify this unit two vector $\hat{\Phi}$,
we distinguish between non-rotating and rotating solutions.
For non-rotating solutions we choose
\begin{equation}
\hat{\Phi} = 
 \left( \begin{array}{c} 
   1 \\ 0 
 \end{array} \right) \ ,
\end{equation}
whereas for rotating solutions we choose
\begin{equation}
\hat{\Phi} = 
 \left( \begin{array}{c} 
   \sin\theta  e^{i \vphi_1} \\ \cos\theta  e^{i \vphi_2} 
 \end{array} \right) \ .
\end{equation}
This choice ensures
single-valuedness of the scalar field, since
\begin{equation}
\Phi(\varphi_1,\varphi_2)=\Phi(2\pi + \varphi_1,2\pi + \varphi_2) \ .
\end{equation}
The specific angular dependence yields a cohomogeneity-1 system.


The choice of appropriate boundary conditions must guarantee
that the solutions
are globally regular and asymptotically flat,
and that they possess a finite energy and finite energy density.
We therefore impose at infinity the boundary conditions
\begin{equation}
f|_{r=\infty}=m|_{r=\infty}=n|_{r=\infty}=1 \ , \ \omega|_{r=\infty}=
\phi|_{r=\infty}=0 \ ,
\label{bc1} \end{equation}
thus the scalar field assumes its vacuum value $\Phi=0$.
At the origin, on the other hand, regularity requires
the boundary conditions
\begin{equation}
\partial_r f|_{r=0}=
\partial_r m|_{r=0}=0
\label{bc2} \end{equation}
together with
\begin{equation}
\partial_r \phi|_{r=0}=0 \ 
\label{bc2a} \end{equation}
for the non-rotating solutions, and
\begin{equation}
\partial_r n|_{r=0}=0 \ , \
\omega|_{r=0}= \phi|_{r=0}=0 
\label{bc2b} \end{equation}
for the rotating solutions.

\subsection{Charges}

The mass $M$ and the angular momenta $J_i$, $i=1,2$, can be obtained
from their respective Komar expressions
\begin{equation}
M = \frac{1}{8\pi G} \frac{3}{2} \int_\Sigma R_{\mu\nu} n^\mu \xi^\nu dV \ , 
\label{Mkomar}
\end{equation}
and
\begin{equation}
J_i = -\frac{1}{8\pi G}\int_\Sigma R_{\mu\nu} n^\mu \eta_{i}^\nu dV \ .
\label{Jkomar}
\end{equation}
Here $\Sigma$ denotes an asymptotically flat spacelike hypersurface,
$n^\mu$ is normal to $\Sigma$ with $n_\mu n^\mu =-1$, $dV$ is the natural
volume element on $\Sigma$, and $\xi$ and $\eta_i$ denote the
Killing vector fields. 
Replacing the Ricci tensor by the stress-energy
tensor yields
\begin{equation}
M = \frac{3}{2} \int_\Sigma \left(T_{\mu\nu} -\frac{1}{3}g_{\mu\nu}T_\gamma^\gamma \right)
 n^\mu \xi^\nu dV \ , 
\label{Mstreng}
\end{equation}
and
\begin{equation}
J_i = - \int_\Sigma \left(T_{\mu\nu} -\frac{1}{3}g_{\mu\nu}T_\gamma^\gamma \right)
 n^\mu \eta_{i}^\nu dV \ . 
\label{Jstreng}
\end{equation}

The mass $M$ and the equal angular momenta $J=J_1=J_2$
can also be obtained directly
from the asymptotic expansion for the metric
\begin{equation}
f=1-\frac{8\, G M }{3\pi r^2} + O\left(\frac{1}{r^4}\right) \ , \ \ \
\omega = \frac{4\, G J }{\pi r^3} + O\left(\frac{1}{r^5}\right) \ .
\end{equation}

The conserved charge $Q$ is associated with the scalar field $\Phi$, 
since the Lagrangian density is invariant 
under the global phase transformation
$\Phi \rightarrow \Phi e^{i\alpha}$. This leads to the conserved current
\begin{equation}
j^\mu = -i \left(\Phi^\dagger \partial^\mu \Phi 
                 - \partial^\mu \Phi^\dagger  \Phi \right)
\ , \ \ \ \ j^\mu_{\ ;\mu} = 0 \ ,
\label{current}
\end{equation}
with the conserved charge $Q$
\begin{equation}
Q = -\int j^t \sqrt{-g} d^4x
. \label{Q_def}
\end{equation}

Inserting the Ansatz into the expressions for 
the angular momenta $J_i$
\begin{equation}
J_i = - 2 \pi^2 \int \sqrt{mnf} \frac{m}{f^3}
\left( \omega_s + \frac{\omega}{r} \right) \phi^2 r^3 dr
\end{equation}
and the conserved charge $Q$
\begin{equation}
Q =  4 \pi^2 \int \sqrt{mnf} \frac{m}{f^3}
\left( \omega_s + \frac{\omega}{r} \right) \phi^2 r^3 dr
\end{equation}
we obtain the quantization condition
\begin{equation}
\sum_i |J_i| = 2|J| = Q \ .
\label{quant}
\end{equation}

\section{Solutions in flat spacetime: $Q$-balls}
In flat spacetime, i.e. for $G=0$ the model possesses $Q$-ball solutions.
In this limit the metric functions are trivial $f=m=n=1$, $\omega=0$.

\subsection{$Q$-ball properties}

The mass $M$ of non-rotating $Q$-balls is given by
\begin{eqnarray}
\label{radenergy}
M 
& = & 2 \pi^2 \int_0^\infty T_{tt} r^3 \, d r \ 
=
2 \pi^2 \int_0^\infty  \left[\omega_s^2 \, \phi^2 +
\phi^{\, \prime \, 2} + U(\phi)\right] r^3 \, d r \ ,
\end{eqnarray}
where the prime denotes differentiation with respect to $r$, 
and their charge $Q$ is given by
\begin{eqnarray}
\label{charge}
Q(\omega_s) & = &  4 \pi^2 \, \omega_s \int_0^\infty \phi^2 \, r^3 \, d r \ .
\end{eqnarray}

As in 4 dimensions \cite{Volkov:2002aj},
$Q$-ball solutions exist in the frequency range
\begin{equation}
\omega^2_{\rm min} < \omega^2_s < \omega^2_{\rm max} \ ,
\label{omega_lim}
\end{equation}
where
\begin{equation}
\omega^2_{\rm max} = \frac{1}{2} U''(0) = \lambda b = m_{\rm B}^2
\ , 
\label{omega_max}
\end{equation}
and
\begin{equation}
\omega^2_{\rm min} = \min_\phi\left[U(\phi)/\phi^2\right] = 
                    \lambda \left(b - \frac{a^2}{4}\right) \ .
\label{omega_min}
\end{equation}
Thus the maximal value is determined only by the boson mass $m_{\rm B}$.

For rotating $Q$-balls the mass receives an additional
contribution and becomes
\begin{eqnarray}
\label{radenergy2}
M =
2 \pi^2 \int_0^\infty  \left[\omega_s^2 \, \phi^2 +
 3 \frac{\phi^2}{r^2} +
\phi^{\, \prime \, 2} + U(\phi)\right] r^3 \, d r \ ,
\end{eqnarray}
The charge and the frequency range remain unchanged.

\subsection{Numerical results}

For the numerical calculations we introduce
the compactified radial coordinate
$\bar{r}= r/(1+r)$ \cite{Kleihaus:1996vi}.
We employ a collocation method for boundary-value ordinary
differential equations, equipped with an adaptive mesh selection procedure
\cite{COLSYS}.
Typical mesh sizes include $10^2-10^3$ points.
The solutions have a relative accuracy of $10^{-6}$.

In the calculations of the 5-dimensional solutions
we choose the same set of potential parameters as
employed previously in 4 dimensions
\cite{Volkov:2002aj,Kleihaus:2005me,Kleihaus:2007vk}
\begin{equation}
\lambda = 1 \ , \ \ \ a=2 \ , \ \ \ b=1.1 \ .
\label{param}
\end{equation}
This yields for the boson mass $m_{\rm B}=\sqrt{1.1}$.
We denote the gravitational coupling constant by $\kappa = 8\pi G$.

\begin{figure}[h!]
\begin{center}
\mbox{\hspace{-1.5cm}
\subfigure[][]{
\includegraphics[height=.27\textheight, angle =0]{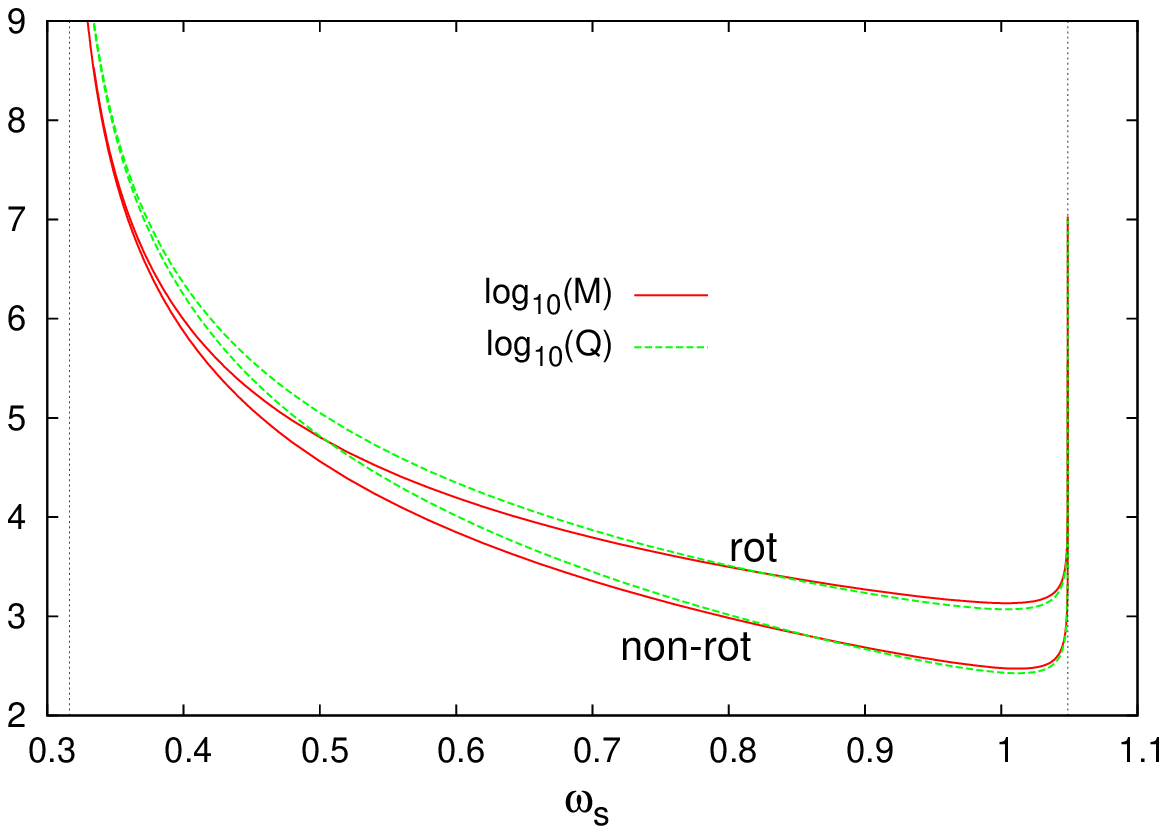}
\label{Q_flat_a}
}
\subfigure[][]{
\includegraphics[height=.27\textheight, angle =0]{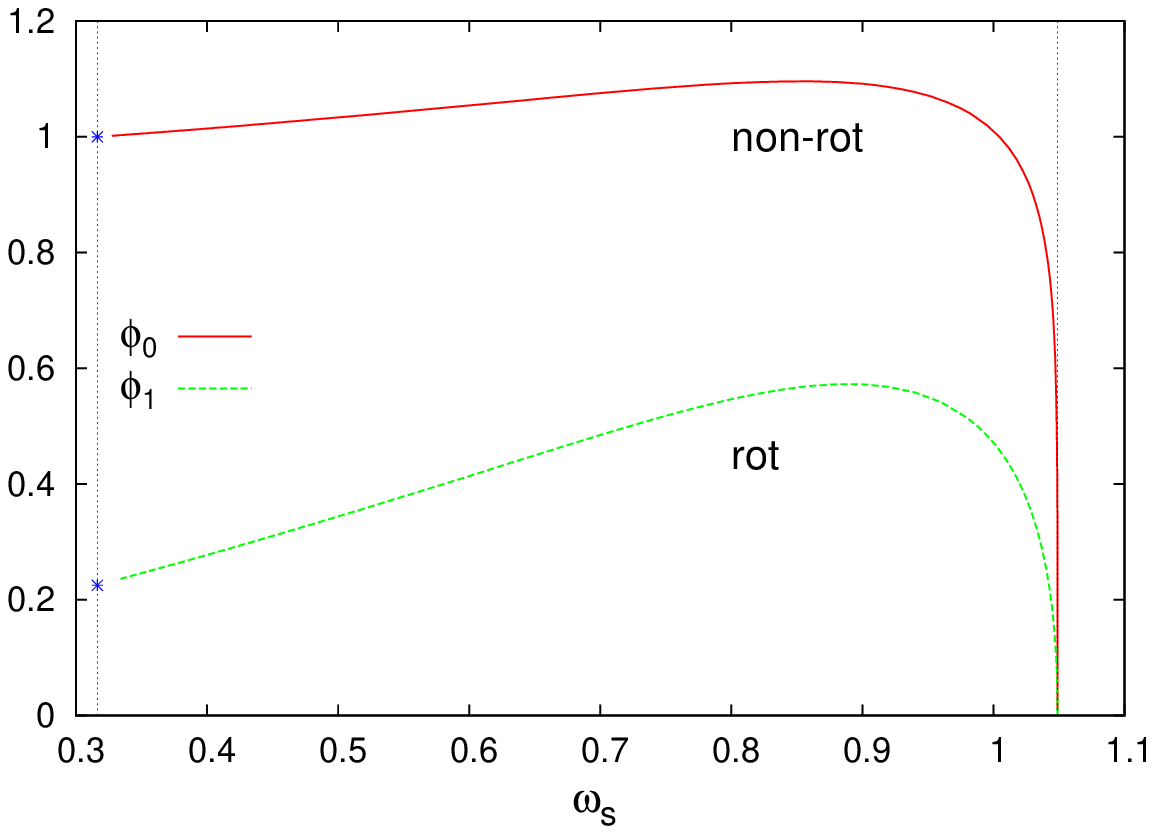}
\label{Q_flat_b}
}
}
\vspace{-0.5cm}
\mbox{\hspace{-1.5cm}
\subfigure[][]{
\includegraphics[height=.27\textheight, angle =0]{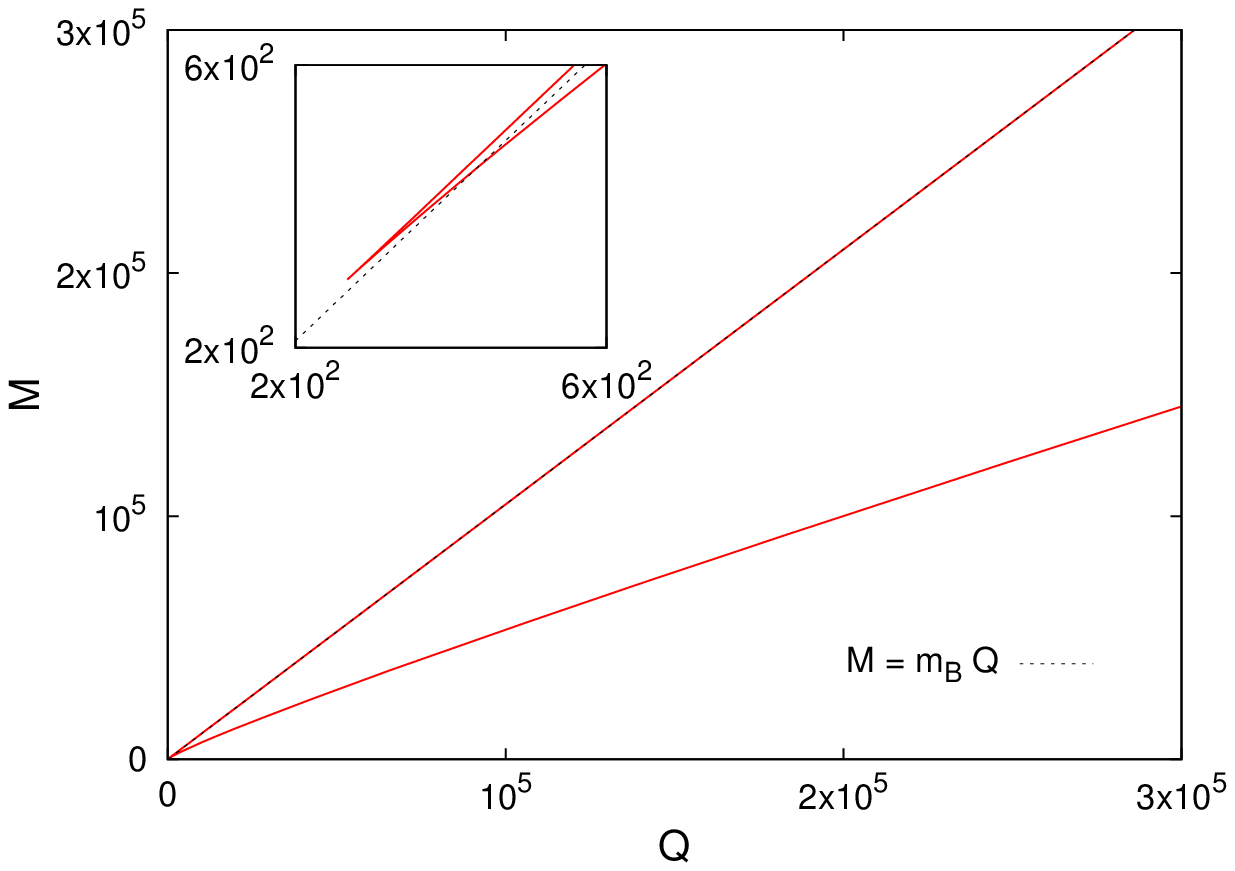}
\label{Q_flat_c}
}
\subfigure[][]{\hspace{-0.5cm}
\includegraphics[height=.27\textheight, angle =0]{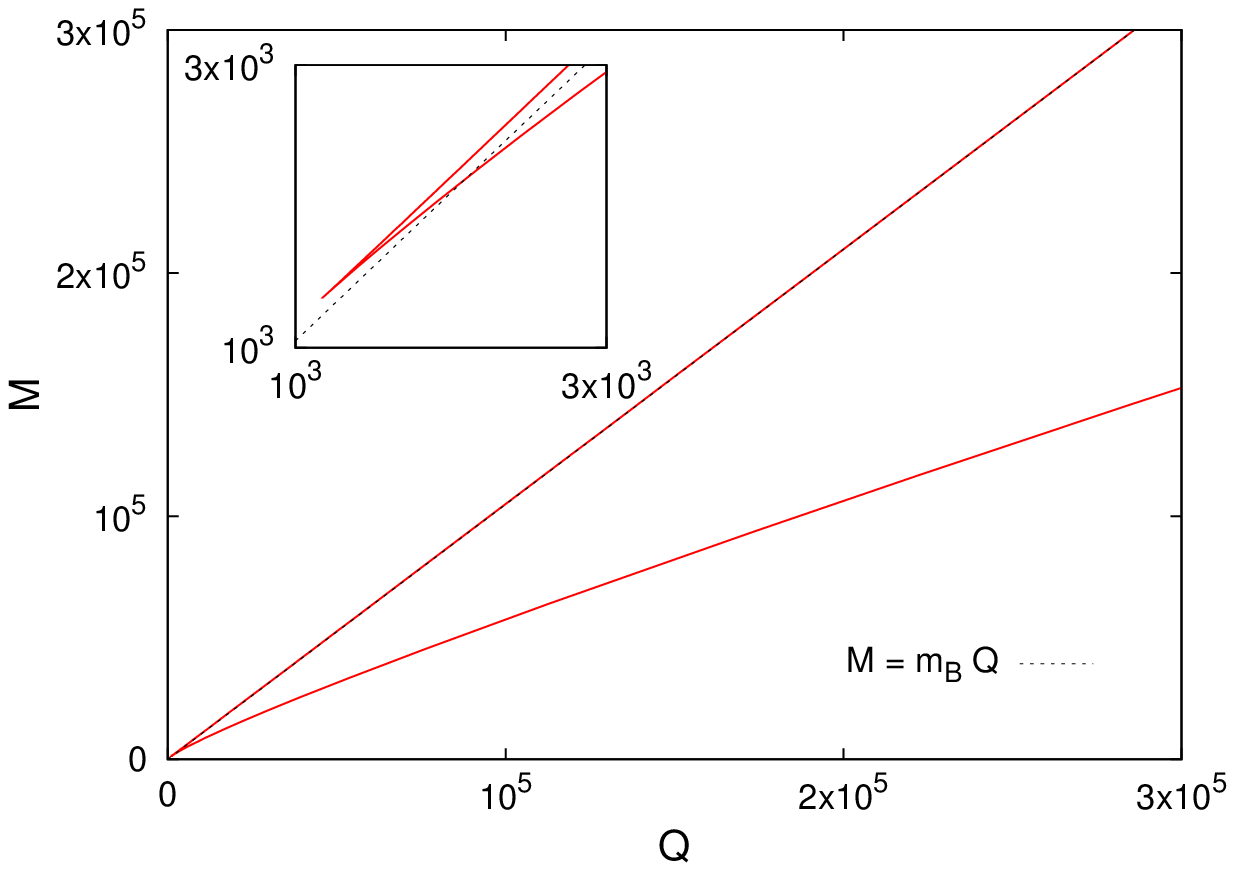}
\label{Q_flat_d}
}
}
\end{center}
\caption{Properties of $Q$-ball solutions are shown:
(a) Mass $M$ and charge $Q$ versus the frequency $\omega_s$;
the vertical lines correspond to the minimal and maximal values of $\omega_s$.
(b) Value of the scalar field function at the origin $\phi_0 \equiv \phi(0)$
for non-rotating $Q$-balls,
and the value of the derivative of the scalar field function at the origin 
$\phi_1 \equiv \partial_r \phi|_{r=0}$
for rotating $Q$-balls. The asterisks indicate the extrapolated limits for 
$\omega_s \to \omega_{\rm min}$.
(c) Mass $M$ versus the charge $Q$ for non-rotating $Q$-balls; 
also shown is the mass of $Q$ free bosons.
(d) Mass $M$ versus the charge $Q$ for rotating $Q$-balls; 
also shown is the mass of $Q$ free bosons.
\label{Q-ball}
}
\end{figure}

In 4 dimensions,
when fixing the value of $\omega_s$ in the allowed range, 
one obtains a sequence of $Q$-ball solutions, 
consisting of the fundamental $Q$-ball and its radial excitations 
\cite{Volkov:2002aj}.
The boson function of the fundamental $Q$-ball has no nodes,
while it has $k$ nodes for the $k$-th radial excitation.
Here we focus only on the fundamental solutions
for $Q$-balls in 5 dimensions.

We exhibit the mass $M$ and charge $Q$ of non-rotating and
rotating $Q$-balls in Fig.~\ref{Q_flat_a} versus the 
frequency $\omega_s$.
At the limits of the interval,
i.e., for $\omega_s \to \omega_{\rm min}$ and
$\omega_s \to \omega_{\rm max}$, both mass and charge diverge.
Inbetween, mass and charge assume their minimal values,
from where they rise monotonically to both sides.
For a given frequency $\omega_s$,
the rotating configurations possess higher mass and charge.
The dependence of the global charges on the frequency
thus is completely analogous for these 5-dimensional $Q$-balls
as for their 4-dimensional counterparts
\cite{Kleihaus:2005me,Kleihaus:2007vk}.

In Fig.~\ref{Q_flat_b}, properties of the scalar field are
exhibited.
For non-rotating $Q$-balls
the value of the scalar field function at the origin $\phi(0)$
is shown versus the frequency,
whereas for rotating $Q$-balls
the value of the derivative of the scalar field function at the origin
$\partial_r \phi|_{r=0}$ is shown.
Both tend to zero at the upper frequency limit, while they assume
finite values at the lower limit.
Inbetween they assume a maximal value.

\subsection{Stability}

To address the stability of these solutions,
we follow the arguments presented in ref.~\cite{Lee:1991ax}.
The stability of the non-rotating solutions
can be read off from Fig.~\ref{Q_flat_c}, where the mass $M$
of the $Q$-balls is exhibited versus their charge $Q$,
together with the mass $M_{\rm f} = m_{\rm B} Q$
of $Q$ free bosons.

We thus observe two branches of $Q$-balls solutions,
a lower branch and an upper branch.
On the lower branch the solutions are classically stable.
As long as $M < M_{\rm f}$, i.e., for most of this lower branch,
they are also quantum mechanically stable.
Shortly after crossing the free mass $M_{\rm f}$, however,
the lower branch ends at the critical point,
where mass and charge assume their minimal values.
The upper branch starts from this critical point,
and approaches the free mass $M_{\rm f}$ asymptotically.
At the critical point the classical stability changes,
and the upper branch has an unstable mode
\cite{Lee:1991ax,Kusmartsev:2008py}.

The situation is analogous for rotating $Q$-balls,
as seen in Fig.~\ref{Q_flat_d}.
Here the critical point, where the classical stability changes,
is encountered for larger values of the mass and charge.
Otherwise, the presence of rotation does not seem to be relevant
for the issue of stability.

\section{Solutions in curved spacetime: boson stars}

We now consider the effect of gravity on these solutions.
The resulting boson stars have many features in common
with their 4-dimensional counterparts. 
But in 5 dimensions also new features arise.
In the following we first discuss the properties of
non-rotating boson stars and then turn to rotating boson stars.

\subsection{Non-rotating boson stars}

\subsubsection{Properties}

\begin{figure}[h!]
\begin{center}
\mbox{\hspace{-1.5cm}
\subfigure[][]{
\includegraphics[height=.27\textheight, angle =0]{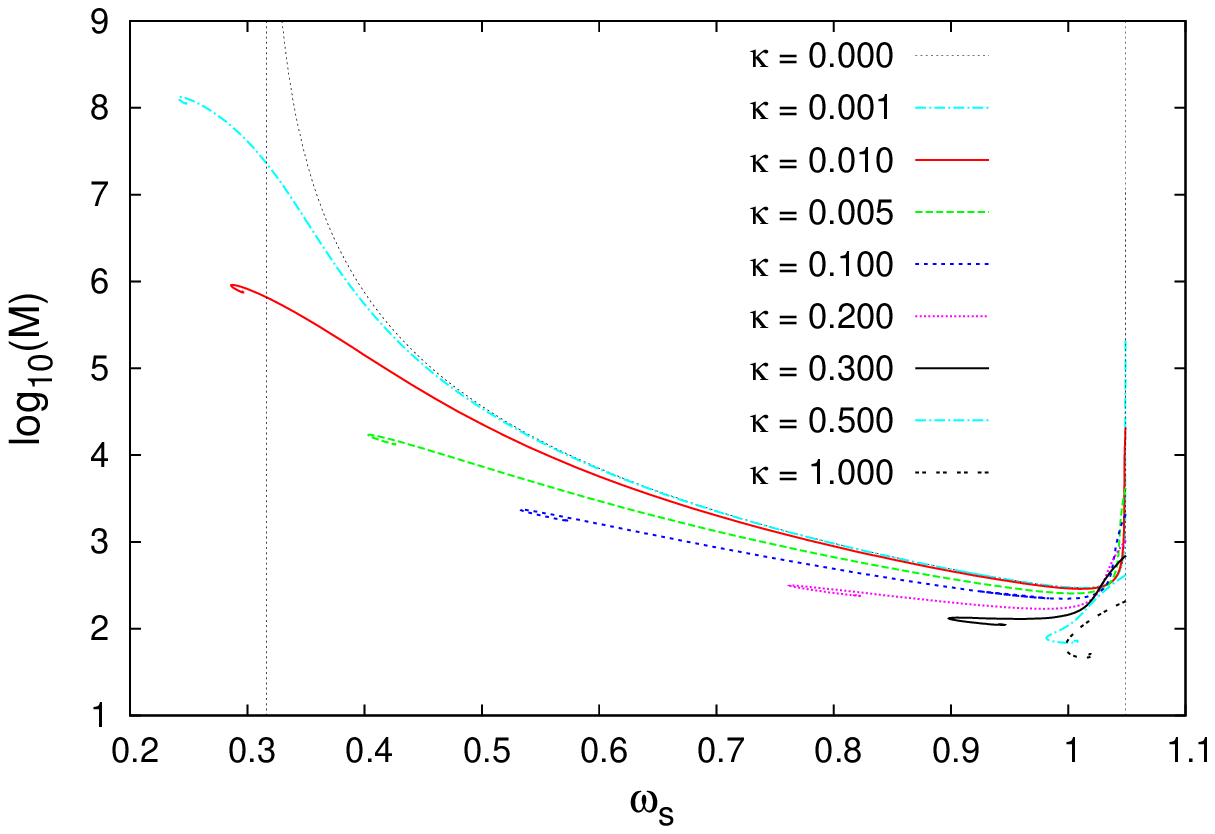}
\label{Stars_a}
}
\subfigure[][]{
\includegraphics[height=.27\textheight, angle =0]{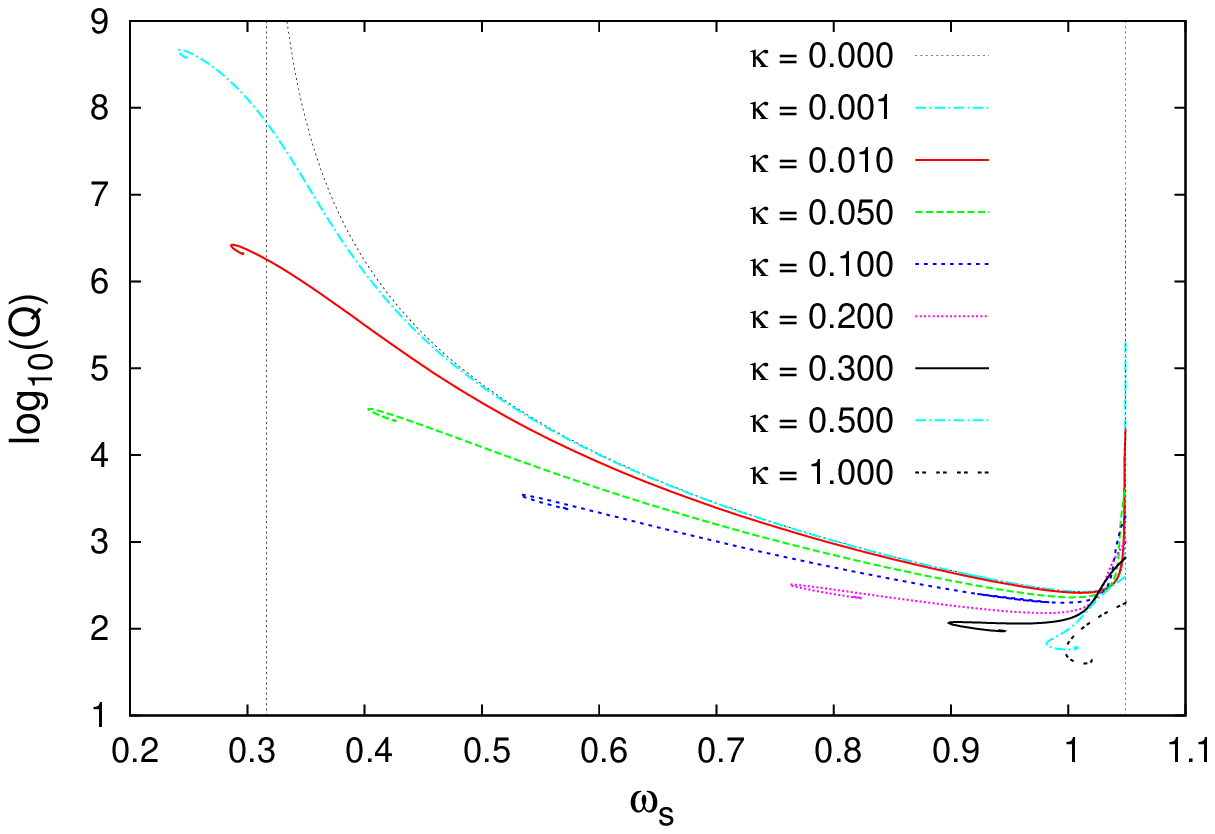}
\label{Stars_b}
}
}
\vspace{-0.5cm}
\mbox{\hspace{-1.5cm}
\subfigure[][]{
\includegraphics[height=.27\textheight, angle =0]{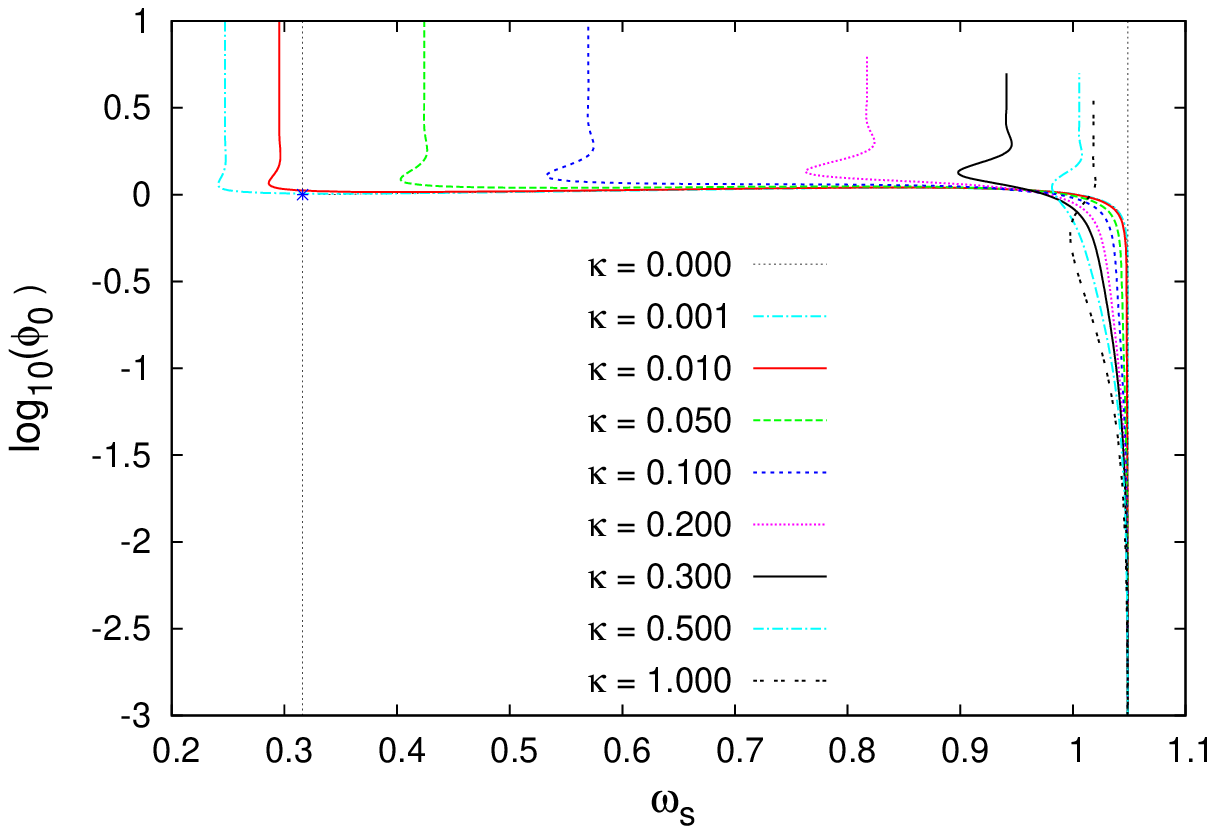}
\label{Stars_c}
}
\subfigure[][]{
\includegraphics[height=.27\textheight, angle =0]{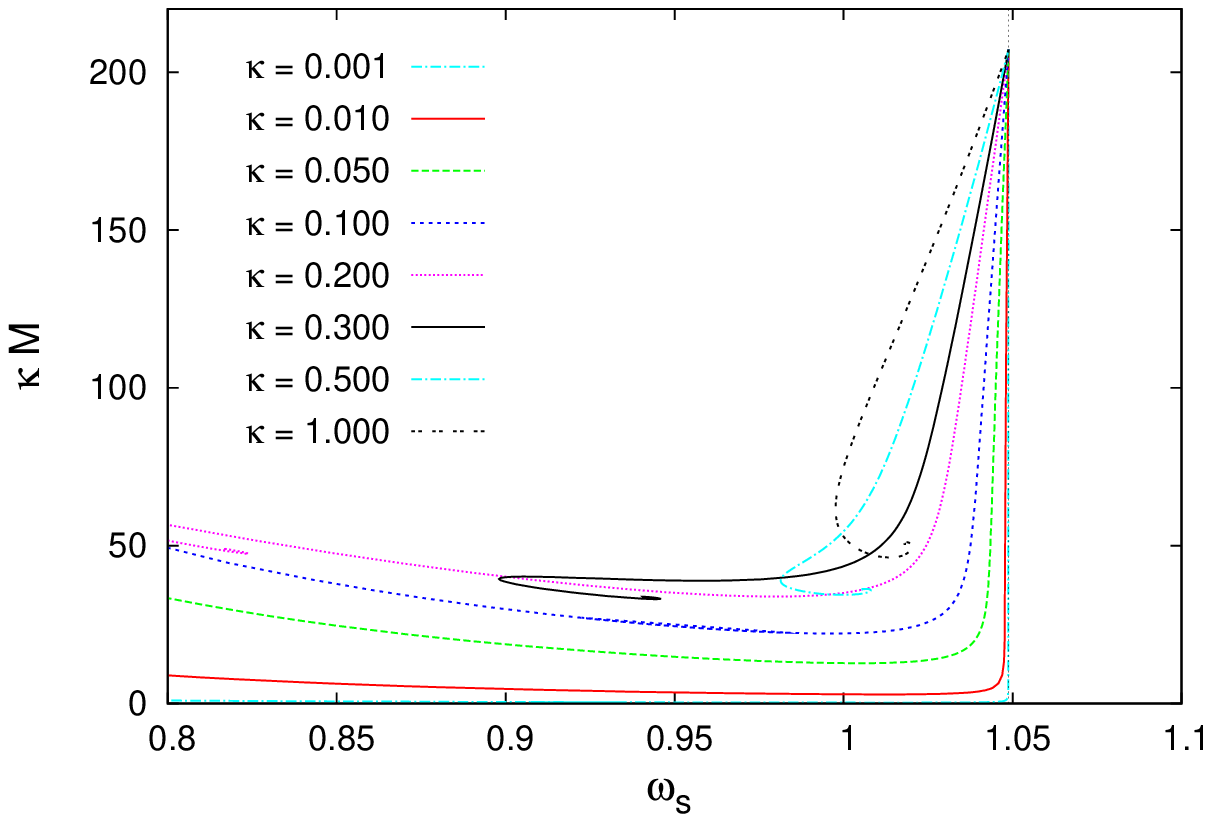}
\label{Stars_d}
}
}
\end{center}
\caption{Properties of the non-rotating boson stars:
(a) Mass $M$ versus rotation frequency $\omega_s$ for several values of the
gravitational coupling constant $\kappa$.
The vertical lines correspond to the flat spacetime minimal 
and universal maximal values of $\omega_s$.
(b) Same as (a) for the charge $Q$.
(c) Same as (a) for the scalar field function $\phi_0 \equiv \phi(0)$
 at the origin. The asterisk indicates the extrapolated limit
$\omega_s \to \omega_{\rm min}$ for $\kappa=0$.
(d) Same as (a) for the scaled mass $\kappa M$.
\label{non-rotBS_prop}
}
\end{figure}

For the boson star solutions in 5 dimensions the frequency 
dependence is in many respects analogous to the 4-dimensional case.
The frequency $\omega_s$
is also bounded from above by
$\omega_{\rm max}$, Eq.~(\ref{omega_max}),
because the scalar field exhibits an exponential fall-off
asymptotically only for $\omega_s < \omega_{\rm max}$.
Moreover, the frequency is also no longer bounded by the flat spacetime
minimal value $\omega_{\rm min}$, Eq.~(\ref{omega_min}),
because the presence of gravity fundamentally changes the dependence of
the mass and charge on the frequency, 
and a spiralling pattern arises at the lower frequency end.

The frequency dependence of the mass $M$ and charge $Q$ 
is exhibited in Figs.~\ref{Stars_a} and \ref{Stars_b}
for non-rotating boson stars
and a sequence of values of the gravitational coupling $\kappa$.
While the presence of a spiral is a genuine property of boson stars,
the location and the size of the spiral depend on
the gravitational coupling strength.
For very small $\kappa$ the spirals have a small extent in
$\omega_s$ and are located beyond $\omega_{\rm min}$, 
thus at smaller values of the frequency
than available in the flat spacetime limit.
With increasing $\kappa$ the spirals then move to
larger frequency values and at the same time cover a larger
range of frequencies.

We exhibit the frequency dependence of
the value of the scalar field at the origin $\phi(0)$
of these boson star solutions
in Fig.\ref{Stars_c}.
The upper endpoints of $\phi(0)$ in the figure correspond to 
values close to the centers of the spirals,
where the numerical procedure was stopped.
Thus these upper endpoints do not signify any physical relevance.
This is in contrast to the universal lower endpoint of $\phi(0)$, 
$\phi(0) \to 0$ for $\omega_s \to \omega_{\rm max}$.

Let us then inspect the behaviour of the boson star
solutions in the limit $\omega_s \to \omega_{\rm max}$
more closely.
Interestingly, in this limit the mass and the charge 
of the 5-dimensional boson stars remain finite.
In 4 dimensions this is not the case, but the
mass and charge tend to zero in this limit.
To obtain further insight into this at first unexpected
behaviour, we consider the scaled mass $\kappa M$
and scaled charge $\kappa Q$.
The scaled mass is exhibited in Fig.~\ref{Stars_d},
and we note that it assumes a universal
value for all boson star solutions in the limit
$\omega_s \to \omega_{\rm max}$,
and so does the scaled charge.

We can understand this behaviour,
when we realize that the metric becomes Minkowskian
in this limit, thus $f=m=1$, 
while the boson field $\phi$ spreads and tends to zero.
At the same time the scaled boson field function 
\begin{equation}
\hat{\phi}(\hat{r})=\phi(\hat{r})/\phi_0\ , \ \ \ \ 
{\rm with} \ \ \ \ \hat{r} =\left(\phi_0\kappa^{\frac{1}{2}}\right)^{\frac{1}{2}} r \ , 
\label{hatr} \ , 
\end{equation}
tends in this limit $\omega_s \to \omega_{\rm max}$
to a universal function for all $\kappa$.

We now express the charge 
in terms of the scaled boson field function $\hat{\phi}$
and the scaled coordinate $\hat{r}$
\begin{equation}
Q = 4 \pi^2 \omega_s \frac{1}{\kappa} \int_0^\infty \hat{\phi}^2 \hat{r}^3 d\hat{r}
\label{Q_lim_omax2} \ , 
\end{equation}
and likewise the mass 
\begin{equation}
M= \frac{2\pi^2}{\kappa}\int_0^\infty 
\left[
\phi_0\kappa^{\frac{1}{2}}\left(\frac{d\hat{\phi}}{d\hat{r}}\right)^2 
+2 \omega^2_s \hat{\phi}^2\right]\hat{r}^3d\hat{r}
\label{K_lim_omax} \ . 
\end{equation}
In the limit $\omega_s \to \omega_{\rm max}$
the value of the scalar field at the origin tends to zero,
$\phi_0 \to 0$, and thus the mass tends to 
\begin{equation}
M
= \frac{4\pi^2}{\kappa}\omega^2_{\rm max}\int_0^\infty \hat{\phi}^2\hat{r}^3d\hat{r}
\label{K_lim_omax2} \ . 
\end{equation}
Since the charge tends to
\begin{equation}
Q = 4 \pi^2 \omega_{\rm max} \frac{1}{\kappa} \int_0^\infty \hat{\phi}^2 \hat{r}^3 d\hat{r}
\label{Q_lim_omax} \ , 
\end{equation}
we obtain the limiting relation between mass and charge
\begin{equation}
M = \omega_{\rm max} Q = m_{\rm B} Q
\label{K_lim_omax3} \ . 
\end{equation}
The explicit limiting values of $\kappa M$ and $\kappa Q$
are determined by the limiting universal function $\hat \phi$.

\begin{figure}[h!]
\begin{center}
\vspace{-1.5cm}
\mbox{\hspace{-1.5cm}
\subfigure[][]{
\includegraphics[height=.27\textheight, angle =0]{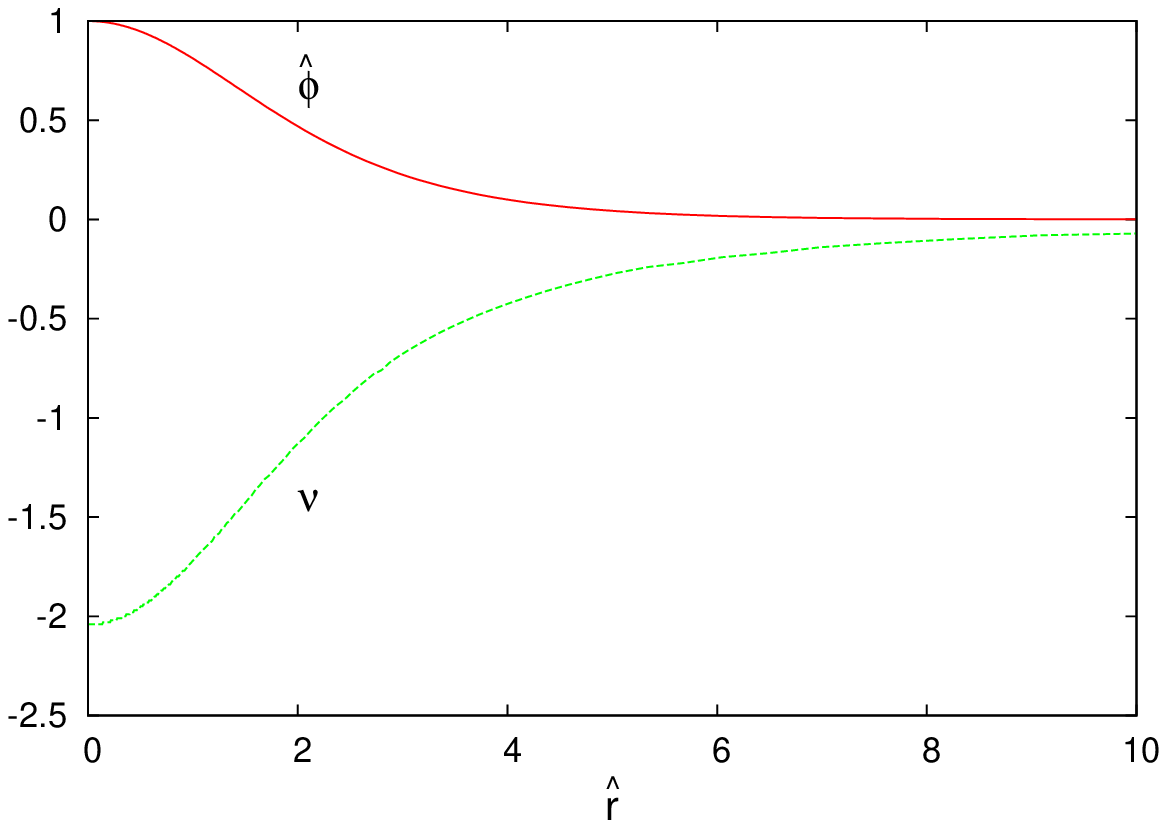}
\label{phinu_vs_rh}
}
\subfigure[][]{
\includegraphics[height=.27\textheight, angle =0]{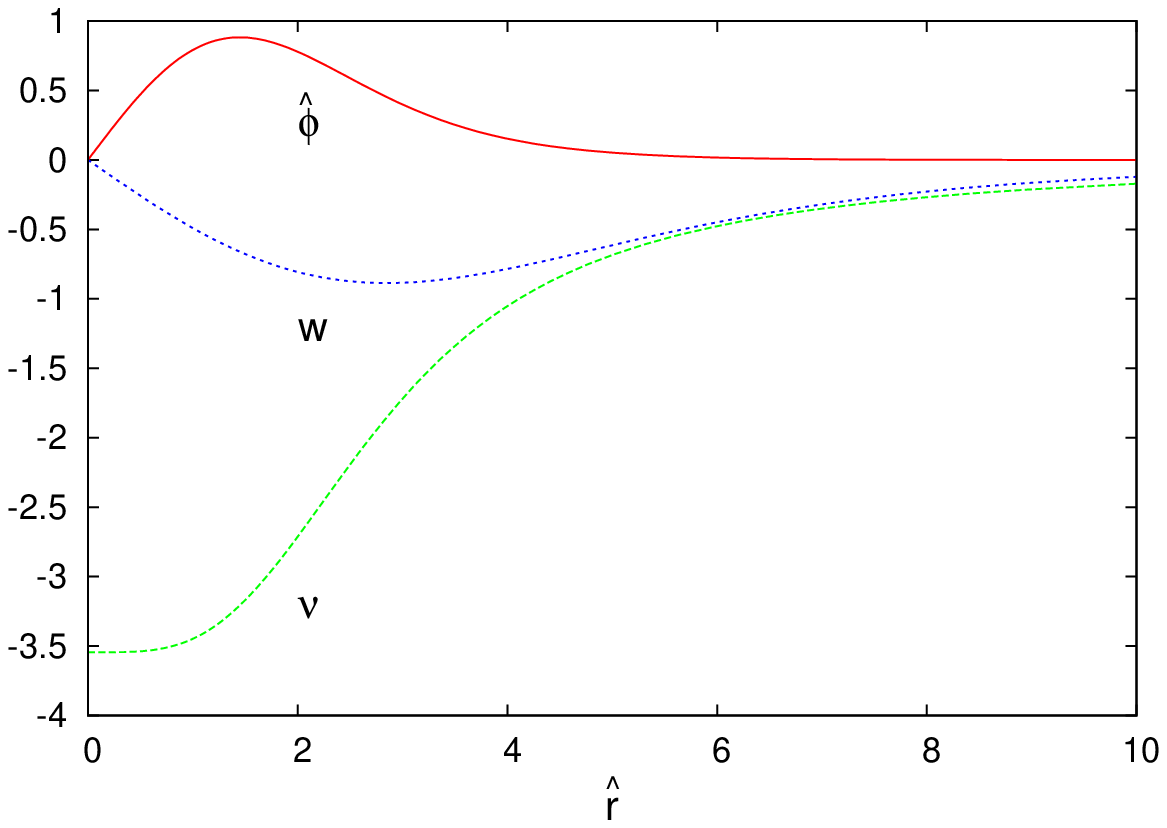}
\label{phinuw_vs_rh}
}
}
\end{center}
\caption{ (a) The function $\hat{\phi}(\hat{r})$ and $\nu(\hat{r})$ are
shown for the non-rotating boson stars. (b) The same for the
rotating boson stars. Also shown is the function $w(\hat{r})$.
\label{phinu_vs_rh_cap}
}
\end{figure}

To determine $\hat \phi$,
we first observe that for all $\kappa$
the ratio $(\omega^2_{\rm max}-\omega^2_s)/\phi_0\kappa^{\frac{1}{2}}$
assumes the same value $\hat{\omega}^2$, say.
We then make a perturbative expansion for the metric functions
\begin{equation}
f = e^{\delta \nu(\hat{r})} \ , \ \ \ \ m = e^{\delta \mu(\hat{r})} \ ,
\label{pertmet}
\end{equation}
where $\delta$ is small.
From the numerical solutions we find $\delta = \phi_0 \kappa^{\frac{1}{2}}$.
Substitution in the boson field equation
and the Einstein equations yields to
lowest order in $\phi_0$
\begin{equation}
\frac{d}{d\hat{r}}\left(\hat{r}^3\frac{d\hat{\phi}}{d\hat{r}} \right)  
= \hat{r}^3 \hat{\phi} \left(\hat{\omega}^2 + \nu \omega^2_{\rm max} \right) 
 \ , \ \ \ \
\frac{d}{d\hat{r}}\left(\hat{r}^3\frac{d\nu}{d\hat{r}} \right)  
= \frac{8}{3}\hat{r}^3 \hat{\phi}^2 \omega^2_{\rm max} 
\ , \ \ \ \ 
\mu= \nu/2 \ .
\label{pertequ}
\end{equation}
The solution to this set of equations is exhibited in Fig.~\ref{phinu_vs_rh}.

Let us now compare the limit $\omega_s \to \omega_{\rm max}$ 
for the boson star solutions in 4 and 5 spacetime dimensions.
In constrast to 5 dimensions, where charge and mass
assume finite values as $\omega_s$ tends to its maximal value,
they vanish in this limit in 4 dimensions.
Analysing the 4-dimensional solutions we observe 
that the scaling behaviour of the boson field function 
is the same as in 5 dimensions, 
see  Eq.~(\ref{hatr}).
However, the volume integral of the mass and charge in 4 dimensions 
contains a power of $r$ less 
than the volume integral in 5 dimensions. 
Consequently the powers of $\phi_0$ do not cancel,
\begin{equation}
Q^{(d=4)} = 8 \pi \omega_{\rm max} \frac{\phi_0^{\frac{1}{2}}}{\kappa^{\frac{3}{4}}} 
                \int_0^\infty \hat{\phi}^2 \hat{r}^2 d\hat{r}
\label{Qd4_lim_omax} \ , 
\end{equation}
and the charge vanishes as $\phi_0$ tends to zero, 
i.e., in the limit $\omega_s \to \omega_{\rm max}$.

\subsubsection{Stability}

\begin{figure}[p!]
\begin{center}
\vspace{-1.5cm}
\mbox{\hspace{-1.5cm}
\subfigure[][]{
\includegraphics[height=.27\textheight, angle =0]{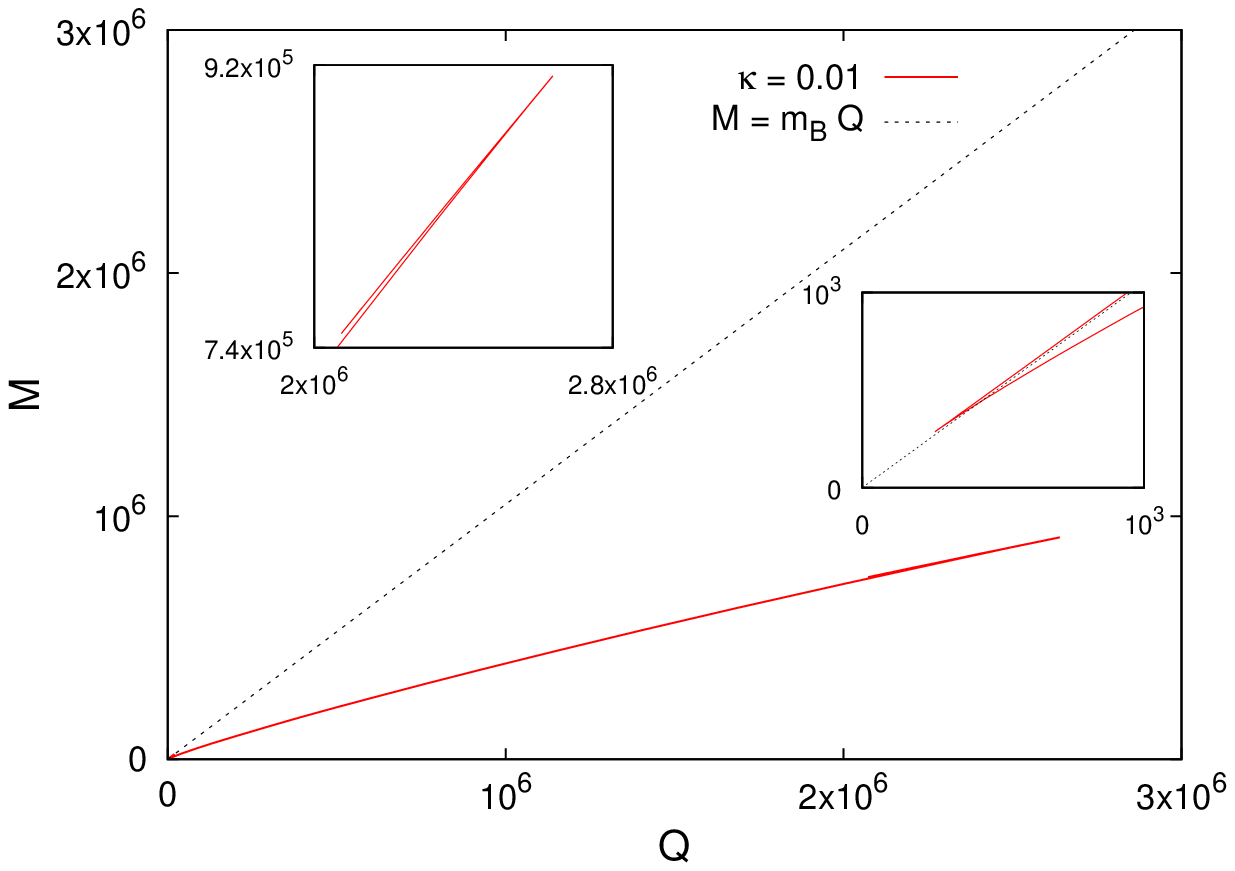}
\label{M_vs_Q_k0.01_nr}
}
\subfigure[][]{
\includegraphics[height=.27\textheight, angle =0]{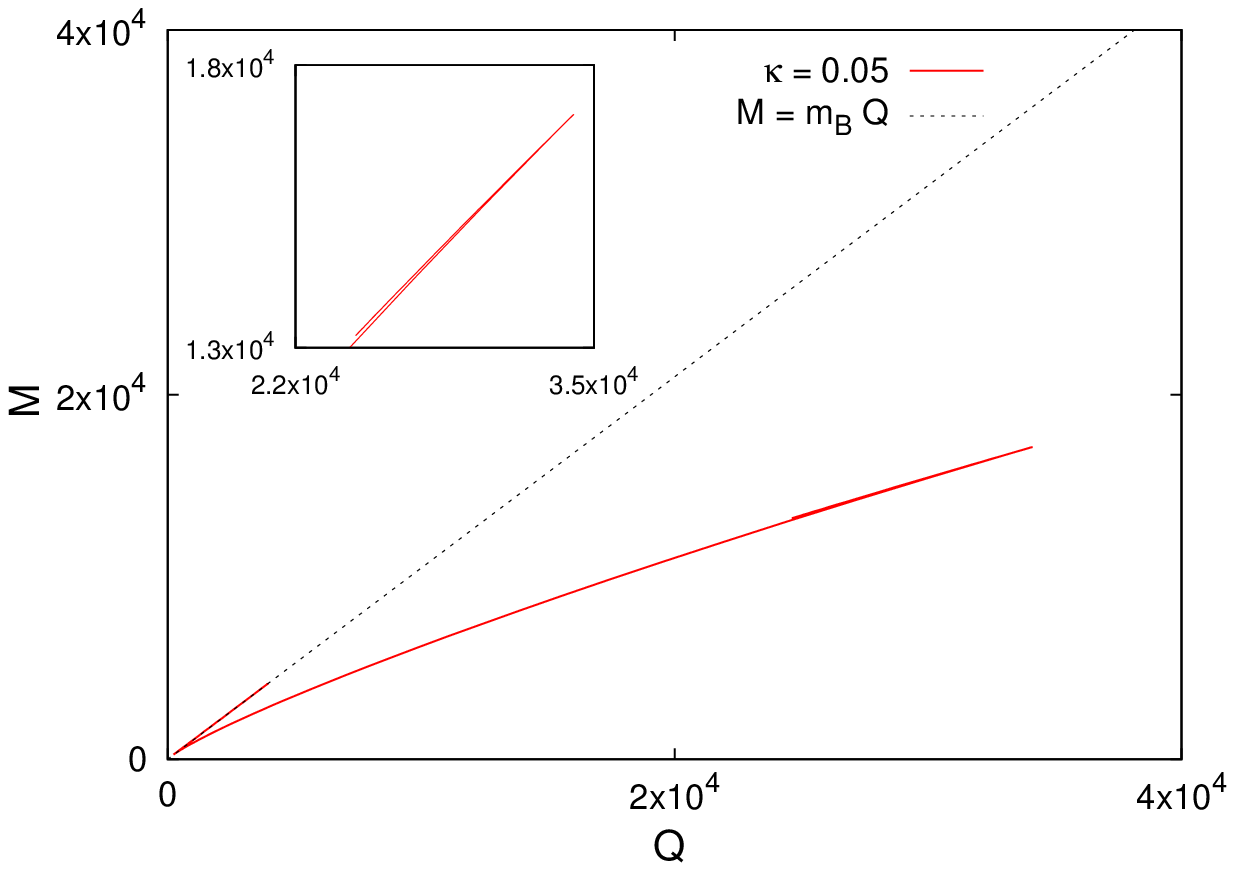}
\label{M_vs_Q_k0.05_nr}
}
}
\vspace{-0.5cm}
\mbox{\hspace{-1.5cm}
\subfigure[][]{
\includegraphics[height=.27\textheight, angle =0]{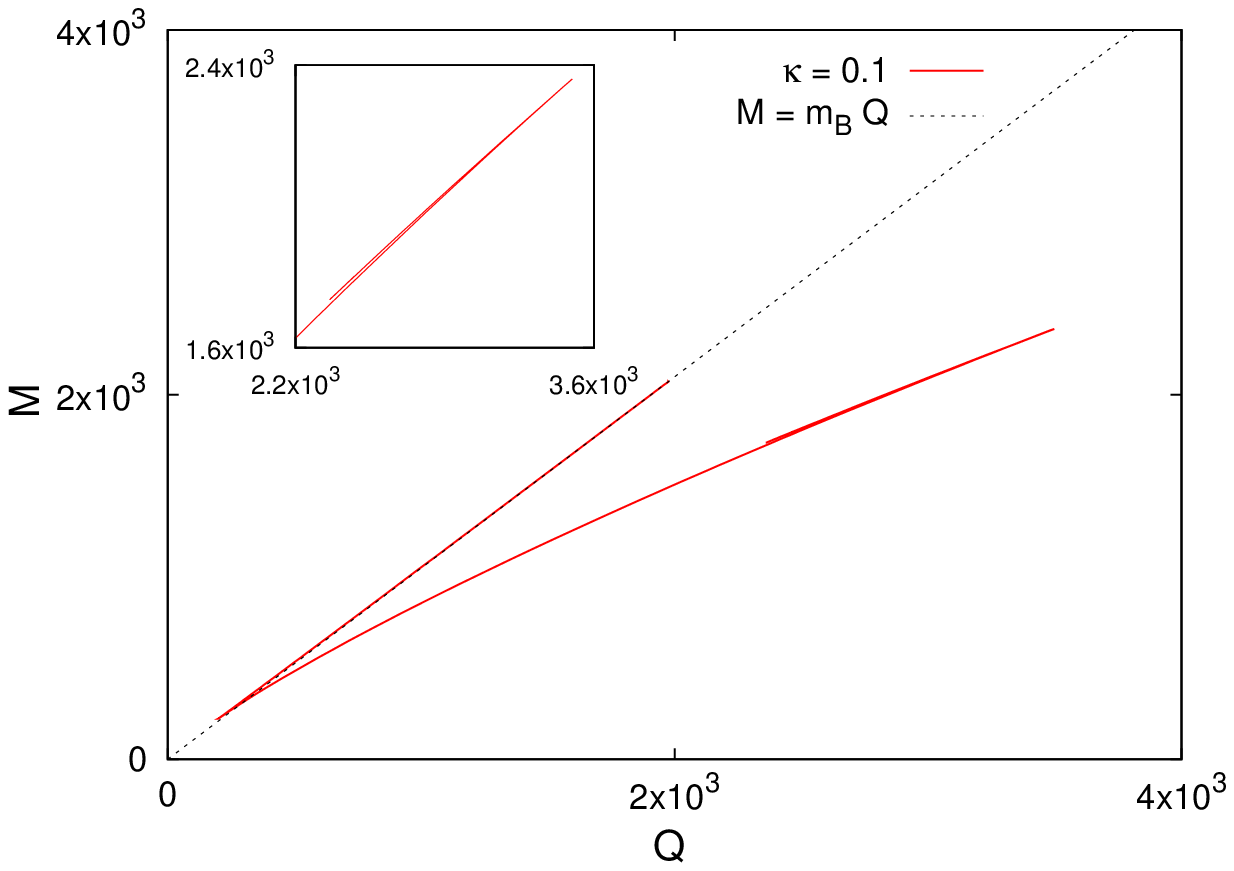}
\label{M_vs_Q_k0.1_nr}
}
\subfigure[][]{
\includegraphics[height=.27\textheight, angle =0]{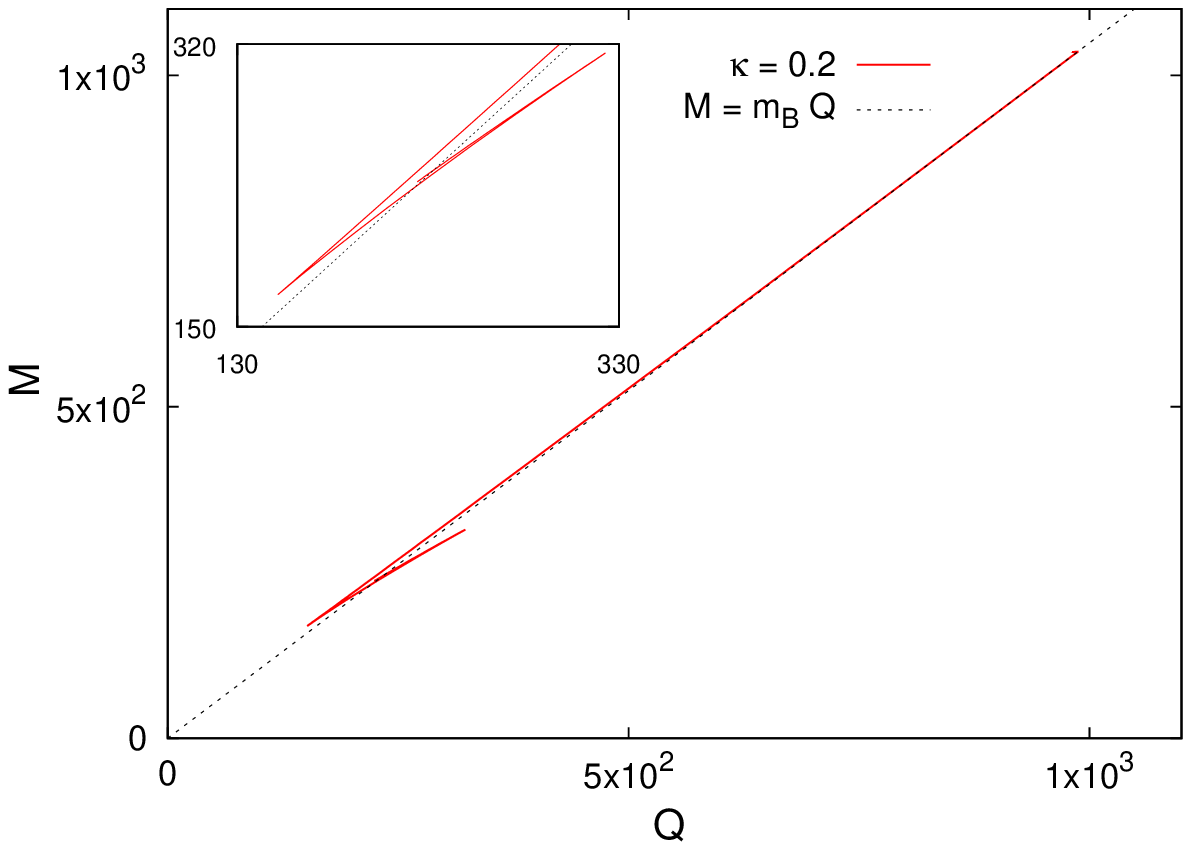}
\label{M_vs_Q_k0.2_nr}
}
}
\vspace{-0.5cm}
\mbox{\hspace{-1.5cm}
\subfigure[][]{
\includegraphics[height=.27\textheight, angle =0]{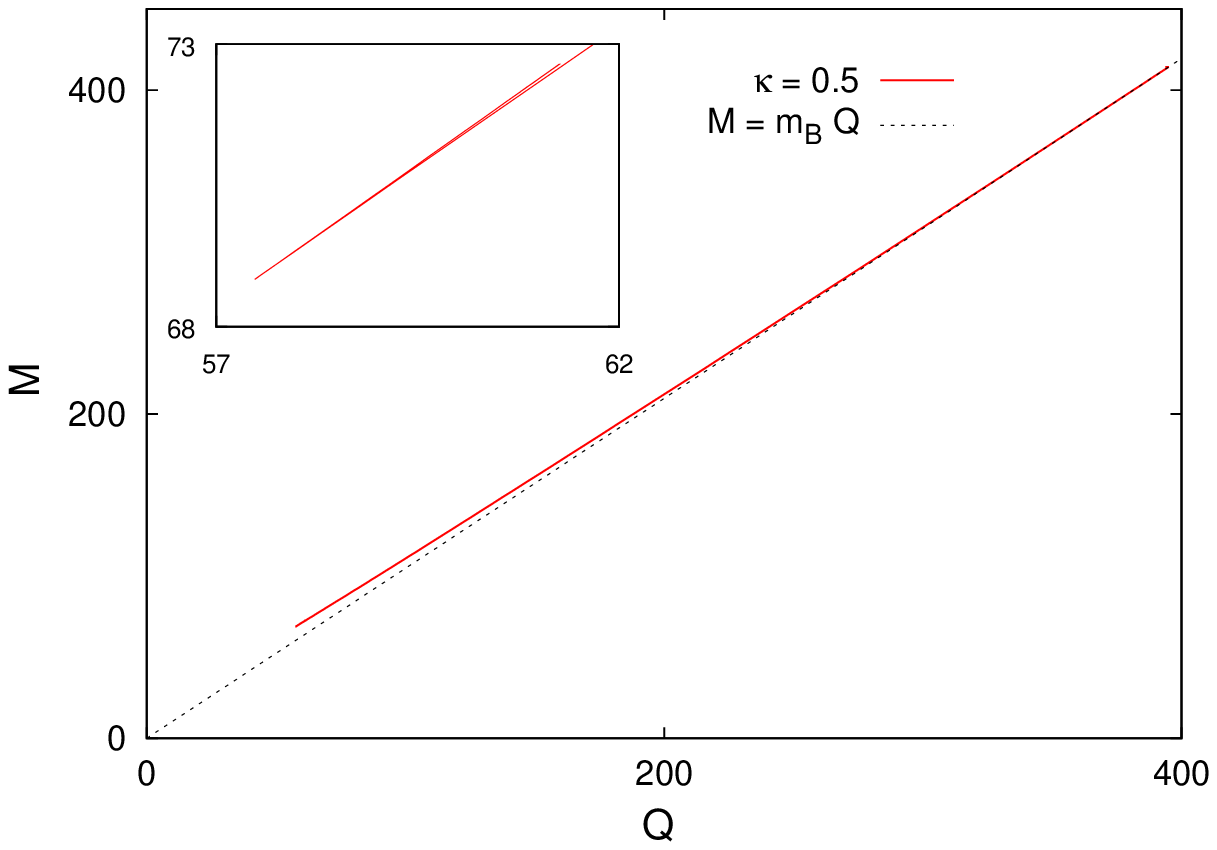}
\label{M_vs_Q_k0.5_nr}
}
\subfigure[][]{
\includegraphics[height=.27\textheight, angle =0]{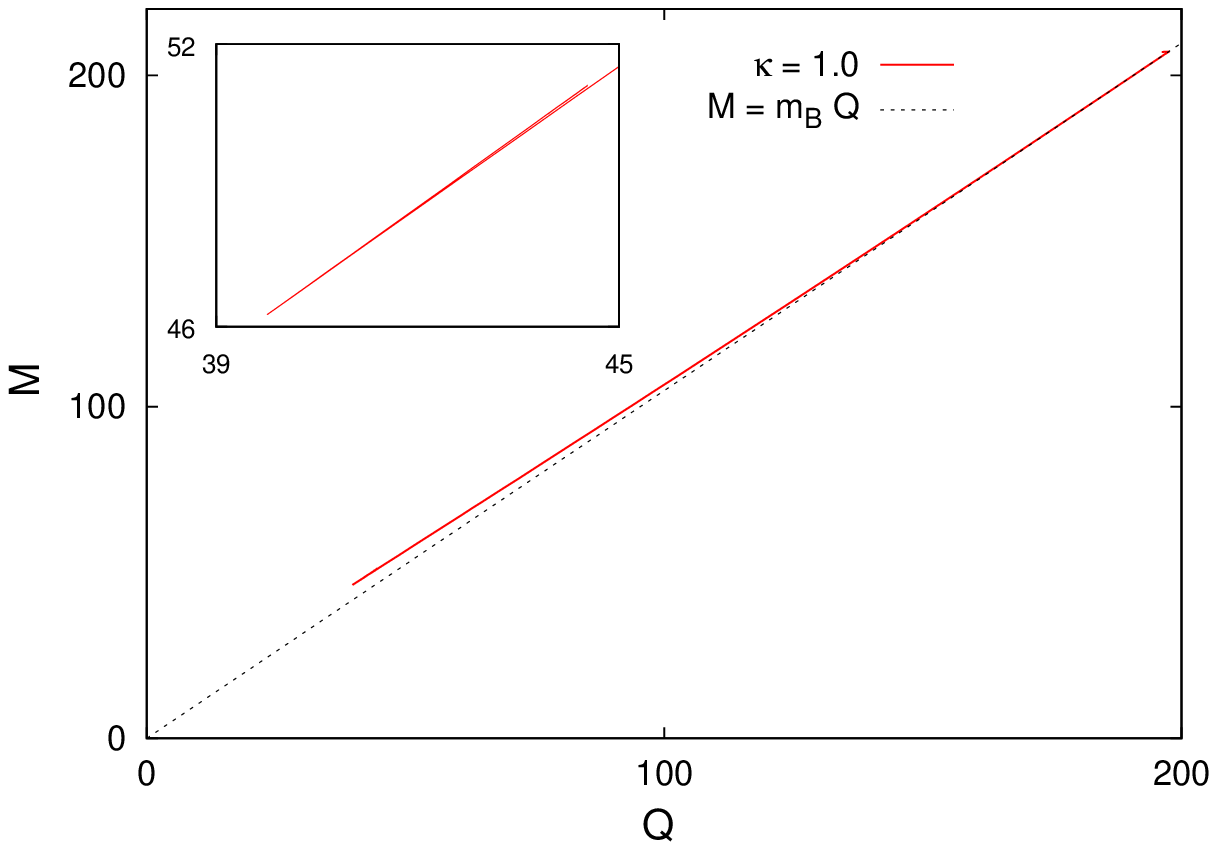}
\label{M_vs_Q_k1.0_nr}
}
}
\end{center}
\caption{Mass $M$ versus charge $Q$ for 
non-rotating boson stars for several values of
the gravitational coupling constant:
(a) $\kappa=0.01$,
(b) $\kappa=0.05$,
(c) $\kappa=0.1$,
(d) $\kappa=0.2$,
(e) $\kappa=0.5$,
(f) $\kappa=1.0$;
also shown is the mass of $Q$ free bosons.
\label{non-rotBS}
}
\end{figure}

Since the mass $M$ and the charge $Q$ have an analogous 
frequency dependence, 
we obtain an intricate cusp structure
when we consider the mass as a function of the charge
as illustrated in Fig.~\ref{non-rotBS}.
As long as gravity is weakly coupled,
e.g., $\kappa \le 0.2$,
the single cusp present in flat spacetime
(associated with the minimal value of the mass and charge)
is retained. 
Here the classically stable branch of boson star solutions ends,
and a classically unstable branch arises.

The boson star solutions on this unstable branch have a mass
that is larger than the mass of $Q$ free bosons.
The branch has a finite extent and ends when 
$\omega_s \to \omega_{\rm max}$,
where $M= M_{\rm f} = m_{\rm B} Q$.
This behaviour is different from 4 dimensions,
where an additional cusp occurs close to 
$\omega_{\rm max}$, because mass and charge tend to zero in the limit.
As seen in Fig.~\ref{non-rotBS},
the relative extent of this classically unstable branch
as compared to the classically stable branch
increases as the coupling $\kappa$ increases.
Being barely noticable for $\kappa=0.01$, Fig.~\ref{M_vs_Q_k0.01_nr},
the classically unstable branch is the dominant branch (qua extent),
when $\kappa=0.2$, Fig.~\ref{M_vs_Q_k0.2_nr}.

The inlets in Fig.~\ref{non-rotBS} demonstrate the cusp
structure due to the presence of the spiral.
Whenever mass and charge assume extremal values,
another cusp arises. According to catastrophe theory,
however, at each new cusp another unstable mode arises
\cite{Kusmartsev:2008py}.
Thus the boson stars become increasingly unstable
as the central point of the spiral is approached.

Turning finally to large values of the gravitational coupling,
we observe, that the classically stable branch has
disappeared completely, since mass and charge no longer
exhibit a minimum outside the spiral.
Moreover, all boson star solutions now have masses above
the mass of $Q$ free bosons.
Thus in this range of $\kappa$ only unstable solutions remain.

Since for large $\kappa$ gravity dominates,
the boson self-interaction terms become negligible,
as demonstrated explicitly in 4 dimensions
\cite{Kleihaus:2005me}.
These solutions then correspond to boson stars
with only a mass term. 
In 5 dimensions such solutions were investigated 
in the presence of a negative cosmological constant
\cite{Astefanesei:2003qy}.
Interestingly, the cosmological constant again
allows for a stable branch,
as explicitly demonstrated 
by Astefanesei and Radu \cite{Astefanesei:2003qy}.

\subsection{Rotating boson stars}

Let us now turn to rotating boson stars in 5 dimensions
with equal angular momenta.
Interestingly, their properties are very similar
to those of the non-rotating boson stars.

\subsubsection{Properties}

\begin{figure}[h!]
\begin{center}
\mbox{\hspace{-1.5cm}
\subfigure[][]{
\includegraphics[height=.27\textheight, angle =0]{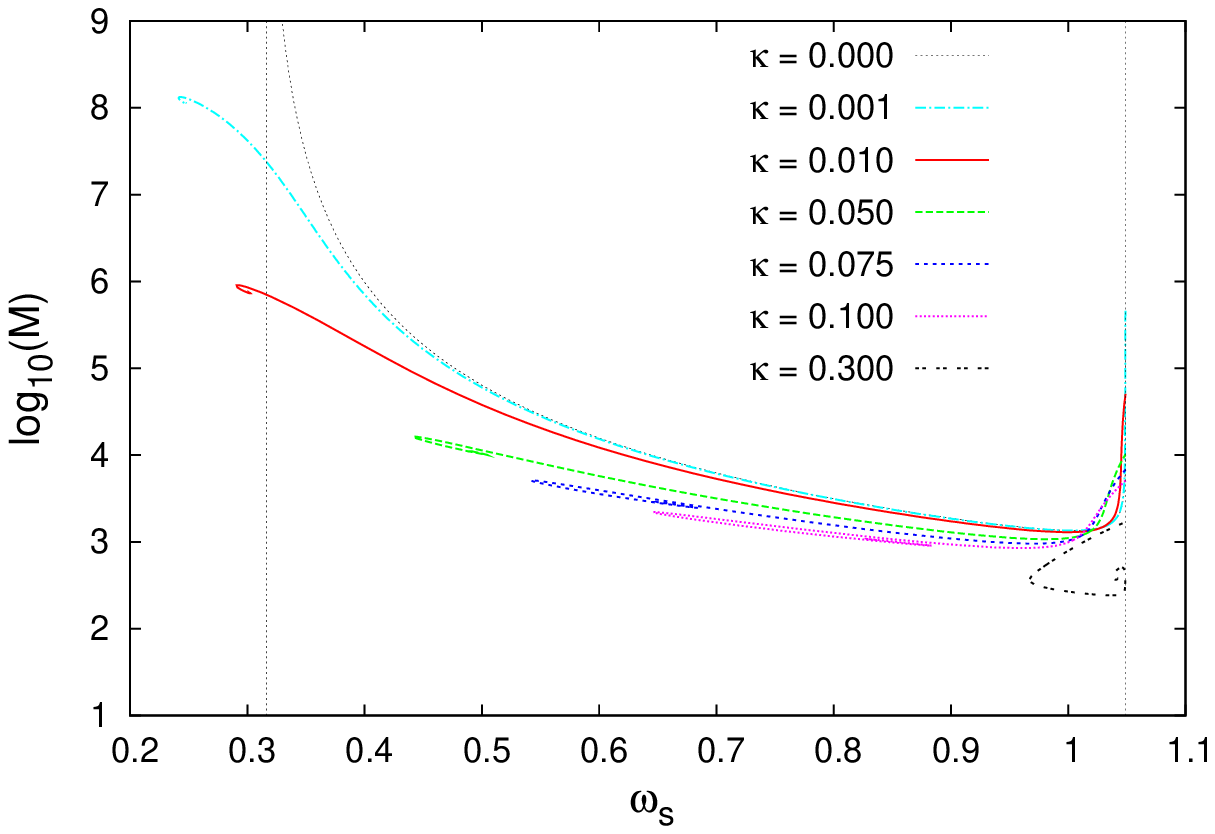}
\label{rot_a}
}
\subfigure[][]{
\includegraphics[height=.27\textheight, angle =0]{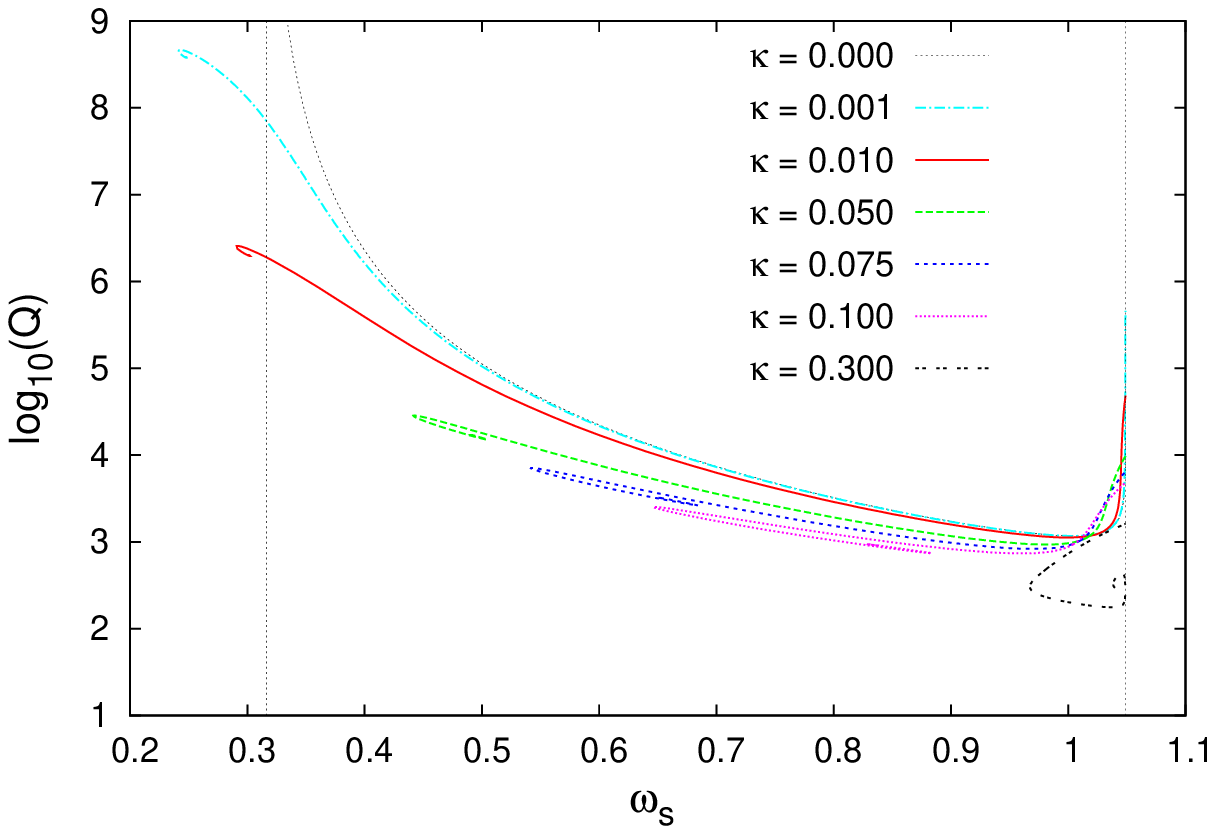}
\label{rot_b}
}
}
\vspace{-0.5cm}
\mbox{\hspace{-1.5cm}
\subfigure[][]{
\includegraphics[height=.27\textheight, angle =0]{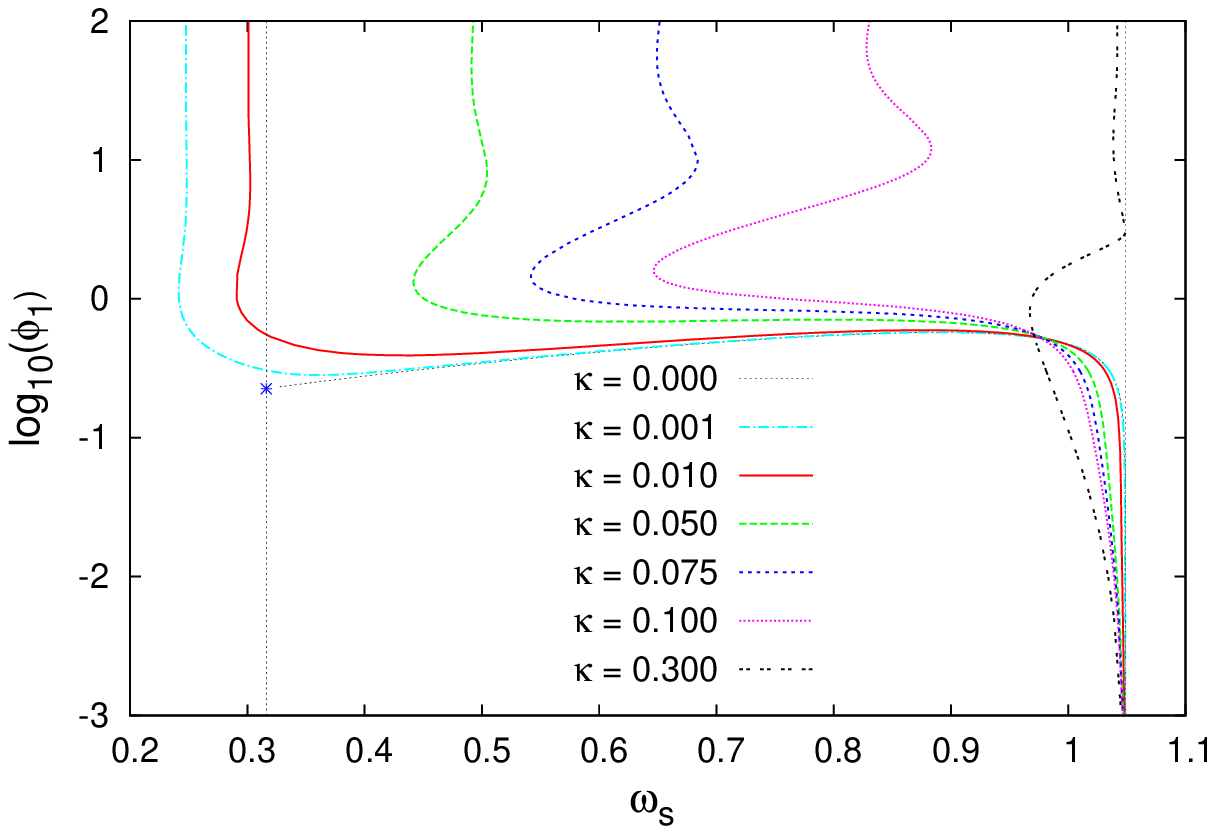}
\label{rot_c}
}
\subfigure[][]{
\includegraphics[height=.27\textheight, angle =0]{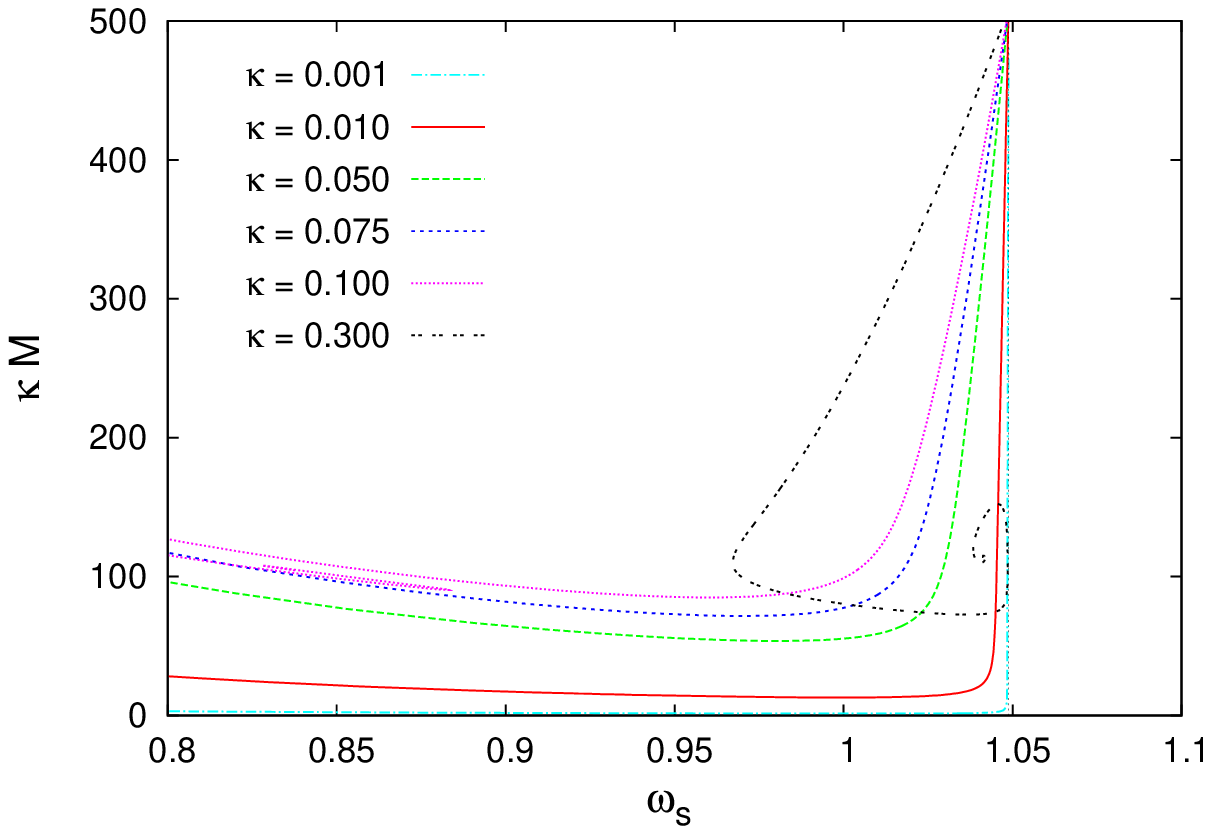}
\label{rot_d}
}
}
\end{center}
\caption{Properties of rotating boson stars:
(a) Mass $M$ versus rotation frequency $\omega_s$ for several values of the
gravitational coupling constant $\kappa$.
The vertical lines correspond to the flat spacetime minimal
and universal maximal values of $\omega_s$.
(b) Same as (a) for the charge $Q$.
(c) Same as (a) for the derivative of the scalar field function 
$\phi_1 \equiv \partial_r \phi|_{r=0}$
 at the origin. The asterisk indicates the extrapolated limit
$\omega_s \to \omega_{\rm min}$ for $\kappa=0$.
(d) Same as (a) for the scaled mass $\kappa M$.
\label{rotBS_prop}
}
\end{figure}

The frequency dependence of the mass $M$ and charge $Q$
of rotating boson stars is in general
rather similar to the case of non-rotating boson stars,
as seen in Figs.~\ref{rot_a} and \ref{rot_b}
for an analogous sequence of values of the gravitational coupling $\kappa$.
The only striking effect of the rotation is that the
spirals elongate substantially for the somewhat
larger values of the coupling strength.
However, this effect is not unexpected,
since we observed it previously also
for rotating boson stars in 4 dimensions
\cite{Kleihaus:2005me,Kleihaus:2007vk}
(representing cohomogeneity-2 solutions).

The frequency dependence of
the derivative of the scalar field function at the origin
$\phi_1\equiv \partial_r \phi|_{r=0}$,
exhibited in Fig.~\ref{rot_c}, 
nicely illustrates these large spirals
by the large oscillations of $\omega_s$
with increasing $\phi_1$.

Let us now again inspect the behaviour of the boson star
solutions in the limit $\omega_s \to \omega_{\rm max}$.
As in the non-rotating case,
the mass and the charge
of the 5-dimensional rotating boson stars remain finite
in this limit.
The scaled mass $\kappa M$ is exhibited in Fig.~\ref{rot_d},
and we observe again a universal
value for the scaled mass and also for the scaled charge
for all boson star solutions in the limit
$\omega_s \to \omega_{\rm max}$,
with the same relation $M=m_{\rm B} Q$,
Eq.~(\ref{K_lim_omax3}).
However, the explicit values of $\kappa M$ and $\kappa Q$
differ from the non-rotating case.

To see where this difference comes from,
we follow the same set of arguments as in the non-rotating case
and introduce a scaled boson field function $\hat \phi$ and
scaled radial coordinate $\hat r$.
In the rotating case we find
\begin{equation}
\hat{\phi}(\hat{r})=
\phi_1^{-1}\left(\phi_1 \kappa^{\frac{1}{2}}\right)^{\frac{1}{3}}\phi(\hat{r})\ , \ \ \ \ 
{\rm with} \ \ \ \ \hat{r} = \left(\phi_1\kappa^{\frac{1}{2}}\right)^{\frac{1}{3}}r  
\label{hatr_rot} \ . 
\end{equation}
The ratio  $(\omega^2_{\rm max}-\omega^2_s)/(\phi_1\kappa^{\frac{1}{2}})^{\frac{2}{3}}$
assumes the same value $\hat{\omega}_r^2$ for all $\kappa$.
For the perturbative expansion for the metric functions
\begin{equation}
f = e^{\delta \nu(\hat{r})} \ , \ \ \ \ m = e^{\delta \mu(\hat{r})} \ , \ \ \ \ 
n = e^{\delta \chi(\hat{r})} \ , \ \ \ \ \omega = \delta_{\omega} w(\hat{r}) \ ,
\label{pertmet_r}
\end{equation}
we find $\delta = \left(\phi_1\kappa^{\frac{1}{2}}\right)^{\frac{2}{3}}$ and 
 $\delta_{\omega} = \phi_1\kappa^{\frac{1}{2}}$.
Substitution in the boson field equation
and the Einstein equations yields to
lowest order in $\phi_1$
\begin{equation}
\frac{d}{d\hat{r}}\left(\hat{r}^3\frac{d\hat{\phi}}{d\hat{r}} \right) 
-3\hat{r}\hat{\phi} 
= \hat{r}^3 \hat{\phi} \left(\hat{\omega}_r^2 + \nu \omega^2_{\rm max} \right) 
 \ , \ \ \ \
\frac{d}{d\hat{r}}\left(\hat{r}^3\frac{d\nu}{d\hat{r}} \right)  
= \frac{8}{3}\hat{r}^3 \hat{\phi}^2 \omega^2_{\rm max} 
\ , \ \ \ \ 
\mu=\chi= \nu/2  \ ,
\label{pertequ_r}
\end{equation}
and 
\begin{equation}
\frac{d}{d\hat{r}}\left(\hat{r}^3\frac{dw}{d\hat{r}} \right) 
-3\hat{r}w 
= 4\omega^2_{\rm max} \hat{r}^2 \hat{\phi}^2 \ .
\label{pertequ_w}
\end{equation}
Thus the limiting solution for $\hat \phi$ is different
in the rotating and non-rotating case,
as seen in Fig.~\ref{phinu_vs_rh_cap}.
To lowest order in $\phi_1$ the mass $M$ and charge $Q$
reduce to the same expressions as in the non-rotating case, 
Eqs.~(\ref{K_lim_omax2}) and (\ref{Q_lim_omax2}), respectively.
Hence again we find
\begin{equation}
M = \omega_{\rm max} Q = m_{\rm B} Q
\end{equation}
in the limit $\omega_s \to \omega_{\rm max}$.

\subsubsection{Stability}

\begin{figure}[p!]
\begin{center}
\vspace{-1.5cm}
\mbox{\hspace{-1.5cm}
\subfigure[][]{
\includegraphics[height=.27\textheight, angle =0]{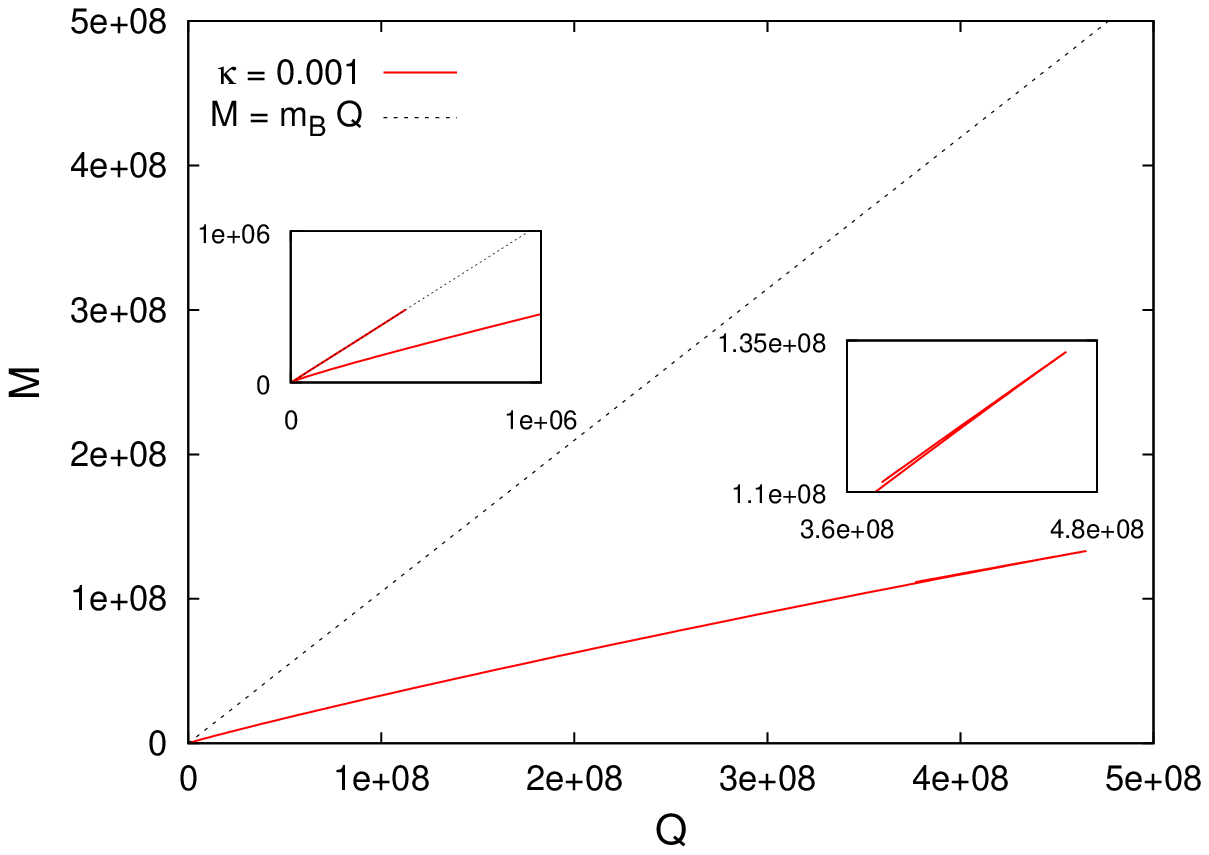}
\label{M_vs_Q_k0.001_r}
}
\subfigure[][]{
\includegraphics[height=.27\textheight, angle =0]{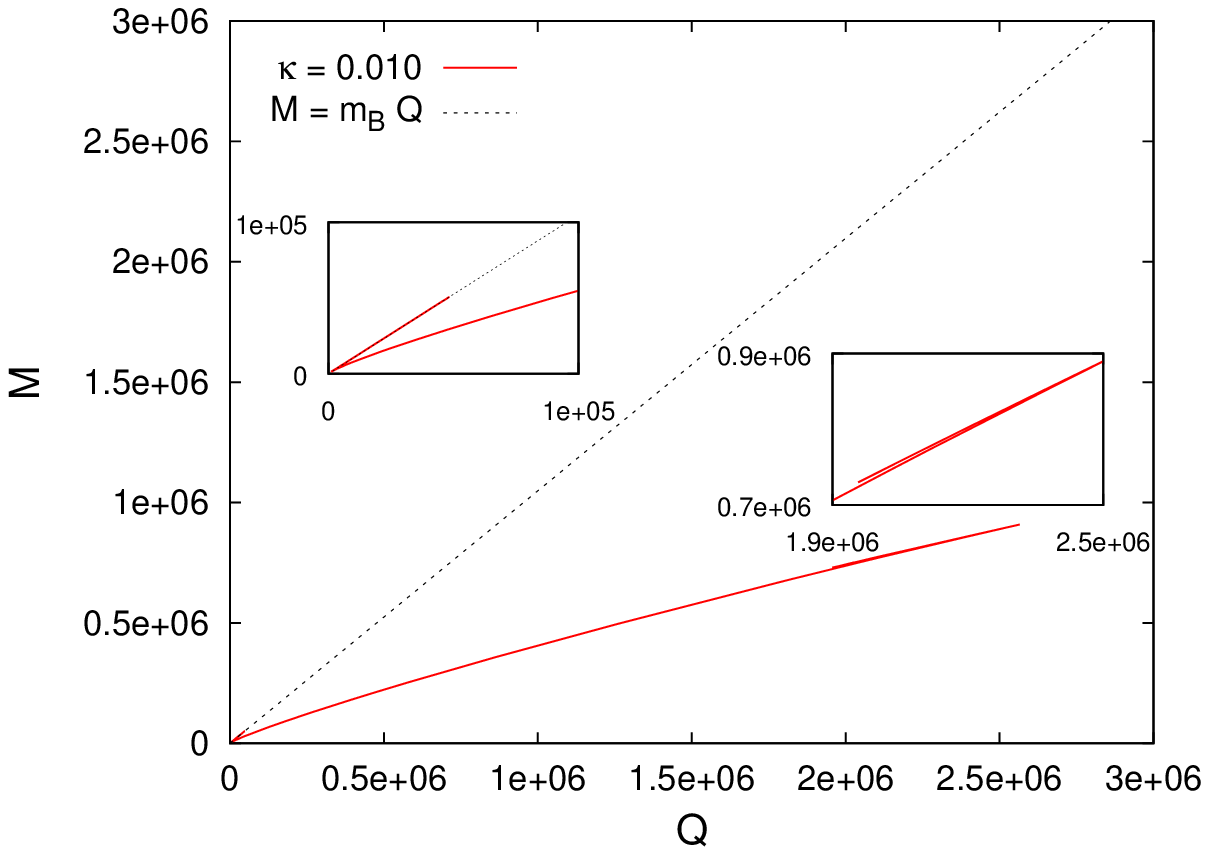}
\label{M_vs_Q_k0.01_r}
}
}
\vspace{-0.5cm}
\mbox{\hspace{-1.5cm}
\subfigure[][]{
\includegraphics[height=.27\textheight, angle =0]{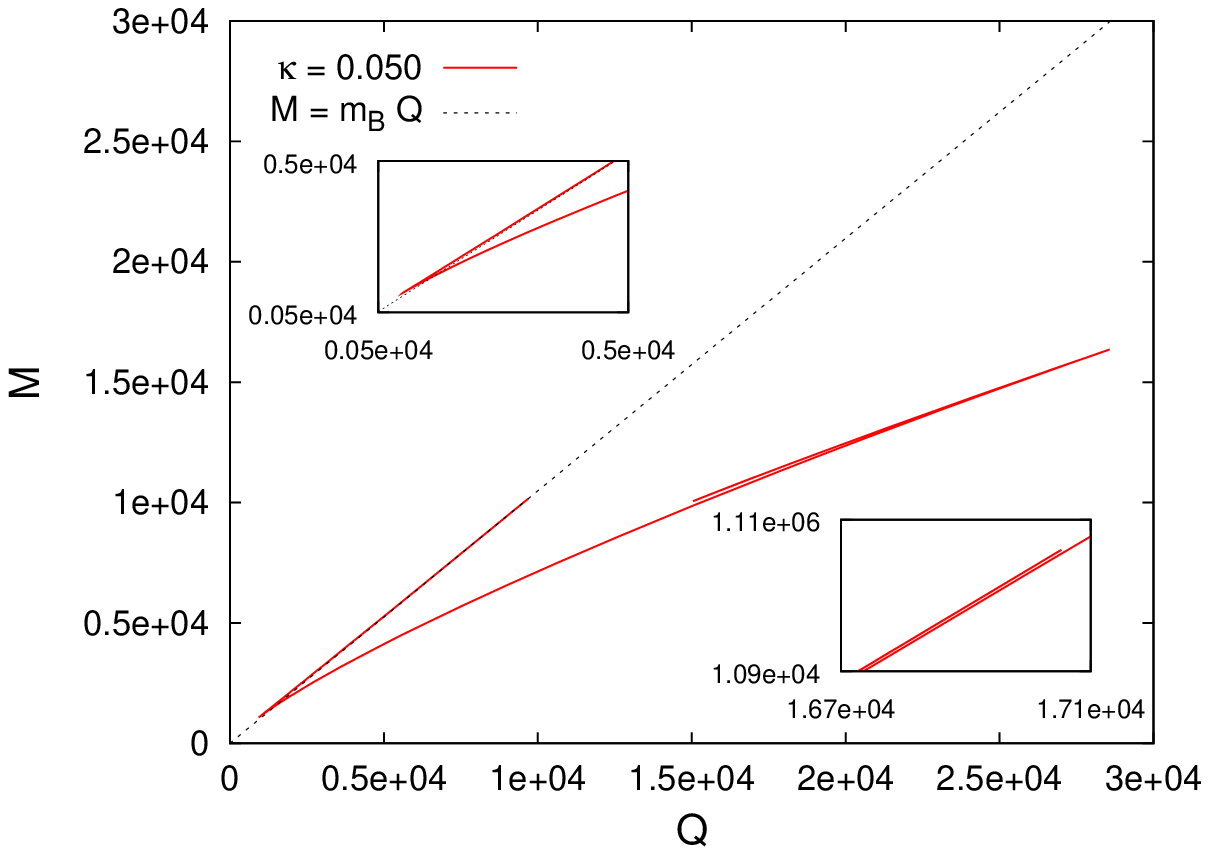}
\label{M_vs_Q_k0.05_r}
}
\subfigure[][]{
\includegraphics[height=.27\textheight, angle =0]{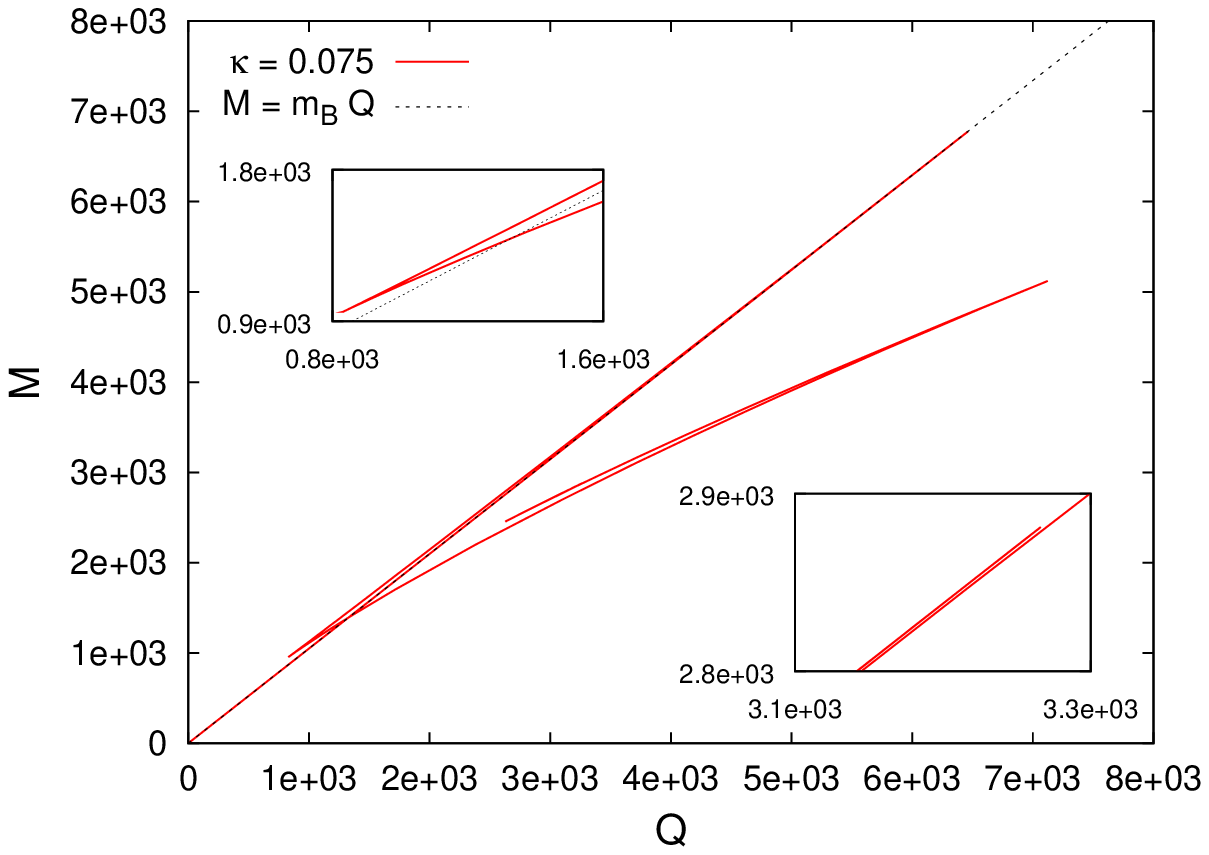}
\label{M_vs_Q_k0.075_r}
}
}
\vspace{-0.5cm}
\mbox{\hspace{-1.5cm}
\subfigure[][]{
\includegraphics[height=.27\textheight, angle =0]{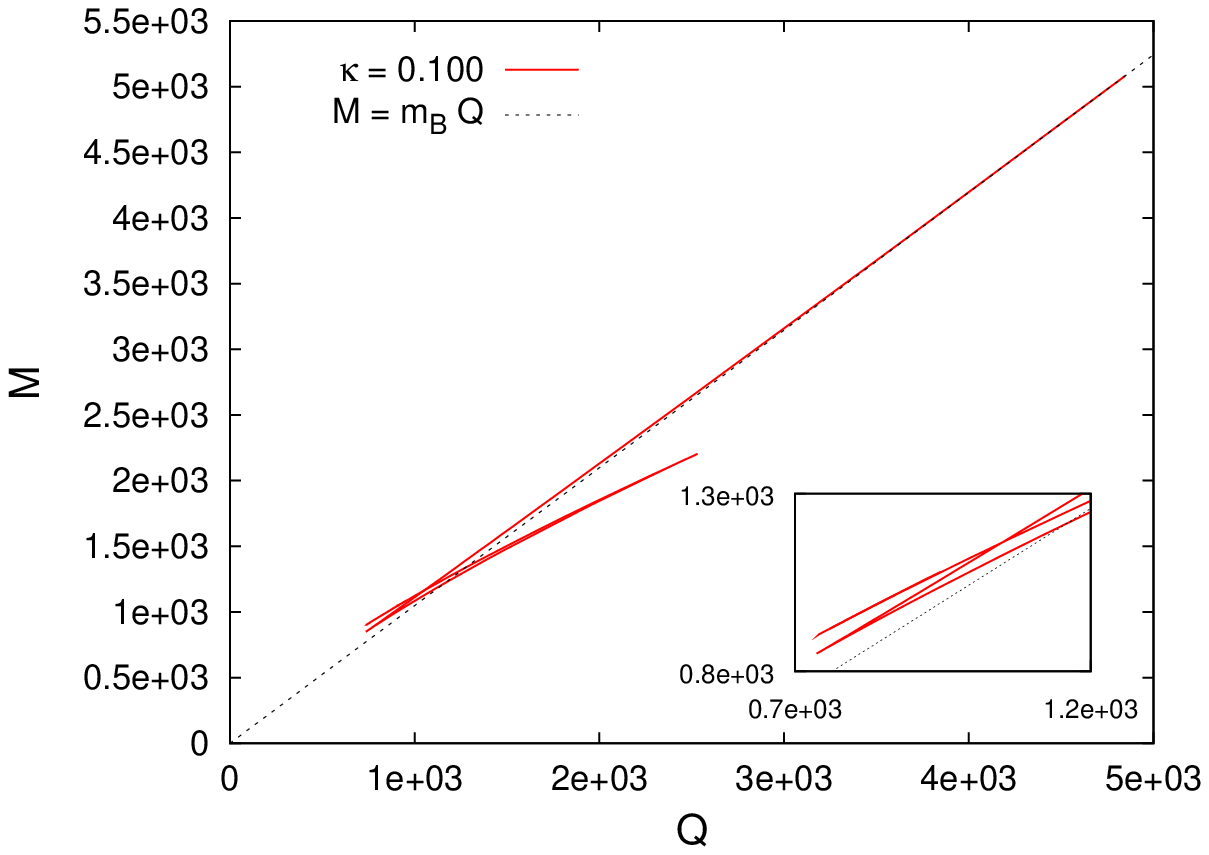}
\label{M_vs_Q_k0.1_r}
}
\subfigure[][]{
\includegraphics[height=.27\textheight, angle =0]{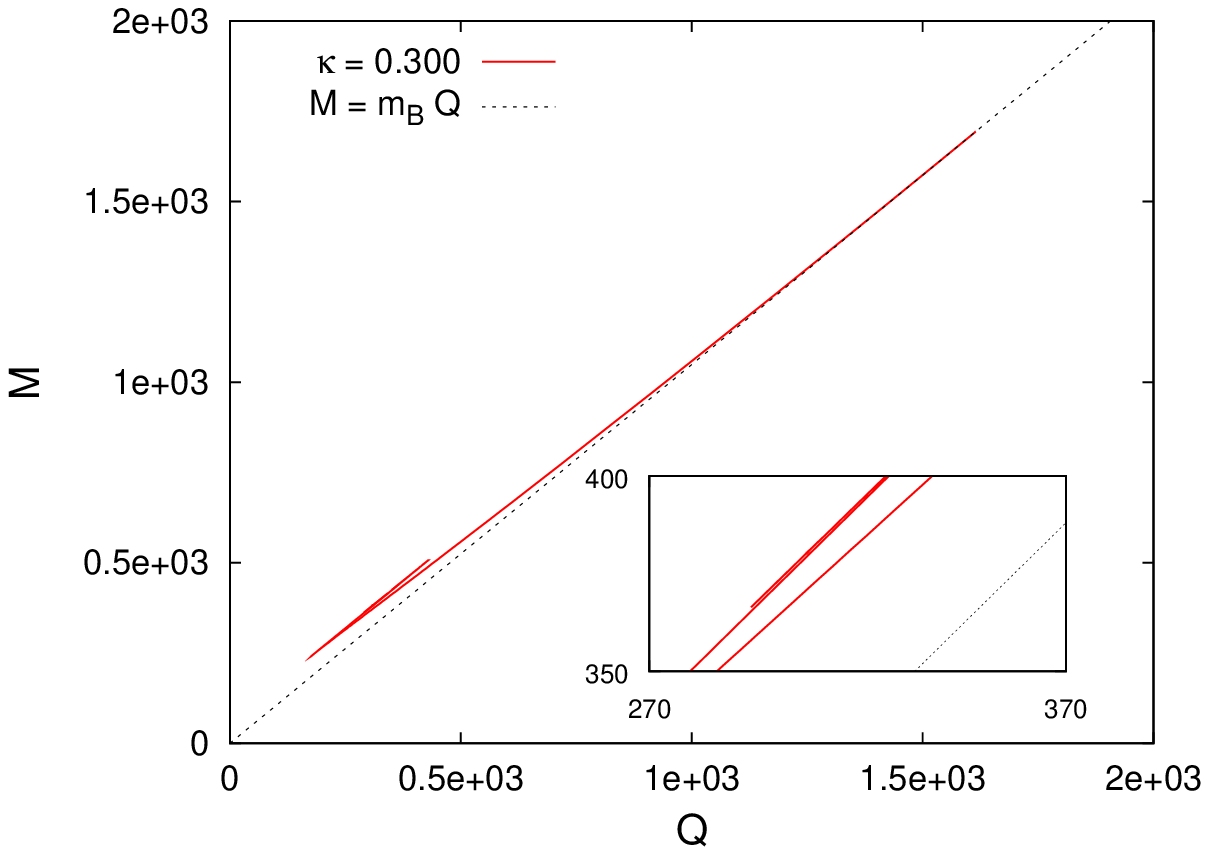}
\label{M_vs_Q_k0.3_r}
}
}
\end{center}
\caption{Mass $M$ versus charge $Q$ 
for rotating boson stars for several values of
the gravitational coupling constant:
(a) $\kappa=0.001$,
(b) $\kappa=0.01$,
(c) $\kappa=0.05$,
(d) $\kappa=0.075$,
(e) $\kappa=0.1$;
(f) $\kappa=0.3$;
also shown is the mass of $Q$ free bosons.
\label{rotBS}
}
\end{figure}

Also the stability analysis for the rotating solutions
is rather analogous to the case of the non-rotating solutions.

When considering the mass as a function of the charge
we obtain the same type of cusp structure,
as seen in Fig.~\ref{rotBS}.
A classically stable branch exists only for weak
gravitational coupling.
With increasing $\kappa$, the relative extent of the
classically stable branch decreases,
whereas the relative extent of the
classically unstable branch increases,
until only unstable solutions remain.

The large size of the spirals make the associated
cusp structure better recognizable for the rotating boson stars.
Still some inlets in Fig.~\ref{rotBS} highlight this cusp
structure due to the presence of the spirals,
where at each new cusp another unstable mode arises
according to catastrophe theory 
\cite{Kusmartsev:2008py}.

\section{Discussion}

We have addressed boson stars and their flat spacetime counterparts, $Q$-balls,
in 5 spacetime dimensions.
Whereas non-rotating solutions are spherically symmetric,
rotating solutions generically possess only axial symmetry,
with two independent planes of rotation and thus
two independent angular momenta.
By employing a complex doublet of boson fields
and choosing a special ansatz, where both angular momenta become equal,
however, the symmetry of the solutions can be enhanced.
The angular dependence then becomes trivial and can be integrated,
simplifying the problem to cohomogeneity-1.

The rotating $Q$-balls and boson stars possess a quantized angular momentum.
In 4 dimensions, for the slowest rotating states,
the quantization condition reads $|J|=Q$.
In 5 dimensions this condition generalizes for the
equal angular momentum solutions studied to
$|J_1|+|J_2|=2|J|=Q$.

As in 4 spacetime dimensions, $Q$-balls and boson stars
exist only in a limited frequency range.
For $Q$-balls we do not observe a qualitative difference in
4 and 5 dimensions.
For boson stars, the same is true for the 
smaller values of the frequency,
where charge and mass exhibit a spiral-like frequency dependence.
In particular, both in 4 and 5 dimensions the effect
of rotation is to considerably elongate the spirals
for not too small gravitational coupling.
However, when the maximal frequency $\omega_{\rm max}$ is approached,
which corresponds to the mass of the bosons,
boson stars show a qualitatively different limiting behaviour.

In 4 dimensions both mass and charge tend to zero
for boson stars as $\omega_s \to \omega_{\rm max}$.
In contrast, in 5 dimensions 
both mass and charge of boson stars tend to finite values
in this limit.
Employing scaled functions and a scaled radial coordinate
we have analyzed this limit explicitly,
by constructing the limiting solutions.
The different behaviour can then be traced back to
the different powers of the radial coordinate in the volume element
in 4 and 5 dimensions.
One may anticipate, that in dimensions greater than 5
the further powers of the radial coordinate 
in the volume element lead to a divergent behaviour of the 
mass and charge. Indeed, we have verified for 
non-rotating boson stars in 6 dimensions, that 
$\phi$ and $\nu$ have the same scaling behaviour 
and thus the mass and charge diverge as $\omega_s \to \omega_{\rm max}$.

The classical stability of $Q$-balls and boson stars
can be analyzed according to catastrophe theory
\cite{Kusmartsev:2008py},
implying a change of classical stability at each cusp encountered,
when the mass is considered as a function of the charge.
While the stability of $Q$-balls does not change
when going from 4 to 5 dimensions,
the stability of boson stars must be reconsidered,
because the cusp structure changes close to $\omega_{\rm max}$.
Thus whereas in 4 dimensions 
there is always a classically stable branch, 
with possible astrophysical relevance,
the boson star solutions in 5 dimensions
may not possess any classically stable branch,
when the gravitational coupling becomes large.

While we have focussed on the fundamental branches
of boson stars, we have also obtained excited (unstable)
branches, where the scalar field function
has up to 6 nodes. A systematic study of these
excited branches is still missing, however.

Another point of interest in the 4-dimensional case
is the possible occurrence of an ergoregion,
since this would imply an instability,
associated with superradiance
\cite{Cardoso:2007az}.
Our previous analysis in 4 dimensions
showed that ergoregions can be present
for boson stars on the classically stable branch.
For boson stars in 5 dimension we observe also ergoregions
in some parameter ranges. Note that in the case of two equal
angular momenta the ergoregion forms a shell.

It is straightforward to extend the present calculations to
non-rotating $Q$-balls and boson stars in more than 5 dimensions.
For the rotating boson stars, however, one needs to restrict 
to odd dimensions and choose all angular momenta equal, 
in order to keep the problem cohomogeneity-1.
Alternatively, allowing for general angular momenta
will increase the numerical complexity tremendously,
since with each new plane of rotation 
the dependence on a further angular coordinate arises.

\vspace{0.5cm}
{\bf Acknowledgement}

\noindent
The authors acknowledge discussion with Eugen Radu.
BK gratefully acknowledges support by the DFG,
and ML by the DLR.

\end{document}